\documentclass[manuscript]{emulateapj}

\usepackage{amssymb}
\usepackage[export]{adjustbox}
\usepackage{amsmath}
\usepackage{booktabs}
\usepackage{hyperref}
\usepackage{multirow}
\usepackage{times}
\PassOptionsToPackage{hyphens}{url}\usepackage{hyperref}
\usepackage[hyphenbreaks]{breakurl}

\newcommand{\degg}{\hbox{$^\circ$}}
\newcommand{\zw}{I\,Zw\,1}
\newcommand{\xmm}{{\it XMM-Newton}}

\newcommand{\ls}
{\mathrel{\hbox{\rlap{\hbox{\lower4pt\hbox{$\sim$}}}\hbox{$<$}}}}
\newcommand{\gs}
{\mathrel{\hbox{\rlap{\hbox{\lower4pt\hbox{$\sim$}}}\hbox{$>$}}}}

\received{}
\begin{document}
\title{A momentum conserving accretion disk wind in the narrow line Seyfert 1, I\,Zwicky\,1.}
\shorttitle{Disk Wind in I\,Zw\,1.}
\shortauthors{Reeves et al.}
\author{J. N. Reeves\altaffilmark{1}, V. Braito\altaffilmark{2,1}}
\altaffiltext{1}{Center for Space Science and Technology, 
University of Maryland Baltimore County, 1000 Hilltop Circle, Baltimore, MD 21250, USA; email jreeves@umbc.edu}
\altaffiltext{2}{INAF, Osservatorio Astronomico di Brera, Via Bianchi 46 I-23807 Merate (LC), Italy}

\begin{abstract}

\zw\ is the prototype optical narrow line Seyfert 1 galaxy. It is also a nearby ($z=0.0611$), luminous QSO, accreting close to the Eddington limit. 
\xmm\ observations of \zw\ in 2015 reveal the presence of a broad and blueshifted P-Cygni iron K profile, as observed through a 
blue-shifted absorption trough at 9\,keV and a broad excess of emission at 7\,keV in the X-ray spectra.
The profile can be well fitted with a wide angle accretion disk wind, with an outflow velocity of at least $-0.25c$. 
In this respect, \zw\ may be an analogous to the prototype fast wind detected in the QSO, PDS\,456, while its overall mass outflow rate is scaled 
down by a factor $\times50$ due to its lower black hole mass. 
The mechanical power of the fast wind in \zw\ is constrained to within $5-15$\% of Eddington, while its momentum rate is of the 
order unity. Upper-limits placed on the energetics of any molecular outflow, from its CO profile measured by {\it IRAM}, appear to rule out the 
presence of a powerful, large scale, energy conserving wind in this AGN. We consider whether \zw\ may be similar to a number of other AGN, such as PDS\,456, 
where the large scale galactic outflow is much weaker than what is anticipated from models of energy conserving feedback.
\end{abstract}










\section{Introduction}


Ultra fast outflows were first detected through observations of blueshifted iron K-shell absorption profiles, as observed in the 
X-ray spectra of Active Galactic Nuclei (AGN). The first known examples of these fast outflows were discovered in the 
luminous quasars, APM\,08279+5255 (\citealt{Chartas02}), PG\,1211+143 (\citealt{Pounds03}) and PDS\,456 (\citealt{Reeves03}). 
Since their initial discovery, a number of high column density ($N_{\rm H}\sim 10^{23}$\,cm$^{-2}$), ultra fast ($\sim0.1c$)
outflows have been found in luminous nearby AGN (\citealt{Tombesi10,Gofford13}). 
These fast winds span a wide velocity range of up to $\sim0.3c$, 
as seen in both PDS\,456 (\citealt{Matzeu17}) and the broad absorption line quasar, APM\,08279+5255 (\citealt{SaezChartas11}).

One of the prototype ultra fast outflows occurs in the nearby ($z=0.184$) QSO, PDS\,456.
By utlizing the hard X-ray bandpass of {\it NuSTAR},  \citet{Nardini15}  revealed a blueshifted P-Cygni profile at iron K in PDS\,456, 
originating from a wide angle accretion disk wind with an outflow velocity of up to $0.3c$. The mechanical power of these winds, like the one observed in PDS\,456, 
can reach a significant fraction of the Eddington limit, which may be more than sufficient to provide the mechanical feedback required by models of black hole and host galaxy co-evolution (\citealt{SilkRees98, Fabian99,DiMatteo05,King03,HopkinsElvis10}). Such black hole winds may play a crucial 
part in the regulating the growth of super massive black holes and the bulges of their host galaxies in luminous QSOs. As black holes grow by accretion, strong nuclear outflows driven by the central AGN can potentially quench this process by shutting off their supply of matter, hereby setting the $M-\sigma$ relation that we see today (\citealt{FerrareseMerritt00,Gebhardt00,Tremaine02}).
In this scenario, gas swept up by the black hole wind within the host galaxy ISM can expand adiabatically, conserving energy in the process and helping to clear matter from the 
galaxy nucleus (\citealt{ZubovasKing12,FaucherQuataert,KingPounds15}). 

AGN molecular gas outflows, observed on kiloparsec scales and with mass outflow rates up to or even exceeding $1000\,{\rm M}_{\odot}$\,yr$^{-1}$, may be the 
signature of this large scale outflowing gas (\citealt{Feruglio10,Maiolino12,Cicone14,Cicone15,Fiore17}). A direct link between the inner, fast black hole winds and the large scale molecular outflows was first made in the type II QSOs, IRAS\,F11119+3257 (\citealt{Tombesi15}) and Mrk\,231 (\citealt{Feruglio15}), via simultaneous detections of both blue-shifted iron K absorption and blue-shifted CO or OH profiles seen in the sub-mm band. 
In these AGN, the momentum rate inferred for the molecular outflow was found to be boosted compared to the inner X-ray wind, consistent the molecular outflow being driven 
by the energy conserving feedback imparted by the initial black hole wind. A third example of a possible energy conserving outflow was then found in the nearby Narrow Line Seyfert~1 (NLS1), IRAS\,17020+4544  (\citealt{Longinotti15,Longinotti18}). 
However, given the limited number of examples in AGN to date, it is vital to explore further examples of fast disk winds, where their impact on larger scale gas can be directly assessed.

The subject of this paper is the nearby ($z=0.0611$) NLS1, I\,Zwicky\,1 (hereafter \zw). 
\zw\ is the prototype optical NLS1, with strong Fe\,\textsc{ii} emission and unusually narrow permitted lines, e.g. 
H$\beta$ FWHM 1240\,km\,s$^{-1}$ (\citealt{Sargent68,OsterbrockPogge85}) and it is also a luminous radio-quiet PG QSO ($M_{B}=-23.5$, \citealt{SchmidtGreen83}). 
It is both a bright and rapidly variable X-ray source (\citealt{Boller96,Gallo04,Wilkins17}). 
Previous \xmm\ observations of \zw\ in 2002 and 2005 have shown the presence of a broad ionized emission line in the iron K band, 
which is centered nearer to 7\,keV rather than from neutral Fe at 6.4\,keV
(\citealt{Porquet04,Gallo04,Gallo07}). This is also consistent with earlier measurements obtained by {\it ASCA} (\citealt{Leighly99,ReevesTurner00}). 

The bolometric luminosity of \zw\ is $L_{\rm bol}\sim3\times10^{45}$\,erg\,s$^{-1}$ (\citealt{Porquet04}), 
which for a black hole mass of $2.8^{+0.6}_{-0.7}\times10^{7}$\,M$_{\odot}$ (\citealt{VestergaardPeterson06}), implies that \zw\ accretes at close to the 
Eddington limit. Such high accretion rate AGN may be prime candidates for driving a fast disk wind. Indeed, in addition to the X-ray winds observed in PG\,1211+143 and 
IRAS\,17020+4544, new examples of ultra fast outflows have also been measured in the NLS1s 1H\,0707-495 (\citealt{Kosec18}) and IRAS\,13224-3809 (\citealt{Parker17,Pinto18}). Furthermore, the possible presence of an ultra fast wind in \zw\ has also been suggested by \citet{Mizumoto19}, who analyzed a small sample of X-ray observations of AGN with existing constraints on the molecular outflow.

Motivated by this, here we present in detail the iron K line profile obtained from all of the \xmm\ observations of \zw\ and we report evidence for a fast ($\sim0.25c$) wide angle wind, similar to the prototype example in PDS\,456. We also compare the energetics derived from the iron K wind with constraints obtained from the sub-mm {\it IRAM} measurements in CO (\citealt{Cicone14}) and demonstrate that the presence of a large scale energy conserving outflow appears to be ruled out in \zw. In Section~2 we describe the \xmm\ observations, while the iron K profile is analyzed in Section~3. In Section~4, the photoionization modeling of the fast outflow is discussed, while in Section~5 the wind profile is fitted with the accretion disk 
wind model of \citet{Sim08,Sim10b}. The energetics of the fast wind are calculated in Section~6 and these are subsequently compared to the molecular outflows observed in other AGN. 

\section{Observations and Data Reduction}

\zw\ was observed four times  with \xmm\  (see Table 1); in 2002 for about $\sim 22$ ksec, in 2005 for $\sim 86 $ ksec and more recently in 2015 in two consecutive orbits for $\sim 141   $\,ks and $\sim 134 $\,ks each.   Each  of the observations was processed and cleaned using the \xmm\ Science Analysis Software (SAS ver. 16.0.0 ) and the latest calibration files available. The resulting spectra were analyzed using the standard software packages (\textsc{xspec} ver. 12.9.1  and \textsc{ftools} ver. 6.22).  Here we concentrate on the  data from the EPIC-pn camera, which once cleaned for any high background returned  the  net exposures listed in Table~1.  Note the net exposures are also corrected for the deadtime of the pn CCD camera.
None of the observations was severely affected by high background with only two minor flares during the last of the observations. All the observations were performed with the thin filter applied and in small window mode, with the exception of the 2005 observation which was operating in the large window mode.
 The source spectra were extracted adopting a circular region  with a radius of 35$''$, while for  the background spectra   were extracted from  two circular regions with  radius of $23''$, $35''$ and $30''$  for the 2002, 2005  and 2015 observations, respectively.
 
 The two observations carried in 2015 were performed in the RGS Multipointing Mode,  which minimizes the flux loss in the RGS at specific wavelengths that are caused by bad pixels or channels. This mode consists of splitting the observation into five different pointings,  which have offsets in  the dispersion direction of 0, $\pm15''$ and $\pm 30 ''$.   Therefore the EPIC observations also consist of five separate pointings, with total exposures ranging from $\sim 20$\,ks to $\sim 28$\,ks each.  We treated each of the five EPIC-pn pointings separately, cleaning the background flares and  extracting the source and background spectra. For each of the pointings we  kept the same radius for the  extraction regions  and  we generated the response matrices and the ancillary response files at the source position using the SAS tasks {\it arfgen} and {\it rmfgen} and the latest calibration available. We subsequently summed the five source and background spectra and combined the response matrices and ancillary response files  with the appropriate weighting, which is proportional to the net exposure time of each of the pointings. 

\section{Spectral Analysis}

We analyzed the two \xmm\ spectra of \zw\ from 2015, using the EPIC-pn CCD detectors, concentrating on the 3--10\,keV band 
in order to study the iron K profile from this AGN in detail.
Note that \zw\ shows rapid X-ray variability on timescales of a few kiloseconds; e.g. see Figure 1, \citet{Wilkins17}. 
However there was no significant variability of the Fe K profile on these timescales and we proceeded to analyze the time averaged spectra 
from each of the two \xmm\ sequences. 
We limited the analysis to the higher energy band, noting that the soft X-ray spectrum below 3\,keV shows a strong soft excess above the power-law continuum, 
while a warm absorber is present in \zw\ at lower energies.
Note that the higher resolution soft X-ray spectra of \zw\ from the Reflection Grating Spectrometer (RGS) on-board \xmm\ have been presented elsewhere. 
\citet{Silva18}  analyzed the RGS from these 2015 observations, while \citet{Costantini07}  analyzed the RGS spectra from the earlier, shorter 2002 and 2005 
\xmm\ observations of \zw. All the soft X-ray spectra show the presence of a two phase warm absorber, with outflow velocities of $
\sim 2000$\,km\,s$^{-1}$. The warm absorber columns densities from the 2015 spectra are of the order $N_{\rm H}=10^{21}$\,cm$^{-2}$ or lower and do not 
impact the hard X-ray spectrum above 3\,keV.

\begin{deluxetable}{lcc}
\tablecaption{Observation log of \zw.}
\tablewidth{250pt}
\tablehead{
\colhead{Start \& End Date} & \colhead{Duration$^{a}$} & \colhead{Net Exposure$^{b}$}}
\startdata
2002-06-22 09:18 $-$ 2002-06-22 15:25  & 21.9 & 18.0\\
2005-07-18 15:22  $-$ 2005-07-18 15:22 &85.5&   57.4\\
2015-01-19 08:56 $-$ 2015-01-21 00:09 &141.2&  89.8\\
2015-01-21 09:20 $-$ 2015-01-22 22:40& 134. 4& 78.3
\enddata
\tablenotetext{b}{Total duration of \xmm\ observation, prior to screening.}
\tablenotetext{b}{Net exposure of EPIC-pn detector after background screening and deadtime correction, in ks.}
\label{tab:obs}
\end{deluxetable}

The 3--10\,keV pn spectra from each of the 2015 observations (OBS\,1 and OBS\,2) are plotted in Figure~1 (panel a), while the background spectra are also shown for comparison. Over the 3-10\,keV band the net source count rates are $0.388\pm0.002$\,cts\,s$^{-1}$ (OBS1\,) and $0.448\pm0.003$\,cts\,s$^{-1}$ (OBS\,2), while the background rates are much lower with $1.53\pm0.04\times10^{-2}$\,cts\,s$^{-1}$ (OBS\,1) and $2.92\pm0.06\times10^{-2}$\,cts\,s$^{-1}$. As a result, the background level has little impact in modeling the spectrum over the Fe K band. Constant energy bins were used for the spectral analysis, sampling the energy resolution of 
the EPIC-pn, which is $\sim160$\,eV (FWHM) at 6\,keV. The spectra have a minimum of 50 source counts per bin, enabling the use of $\chi^2$ minimization in the 
spectral fitting. Note that all error measurements in the subsequent spectral fitting are given at the 90\% confidence level for 1 parameter of interest.

\begin{figure}
\begin{center}
\rotatebox{-90}{\includegraphics[height=8.5cm]{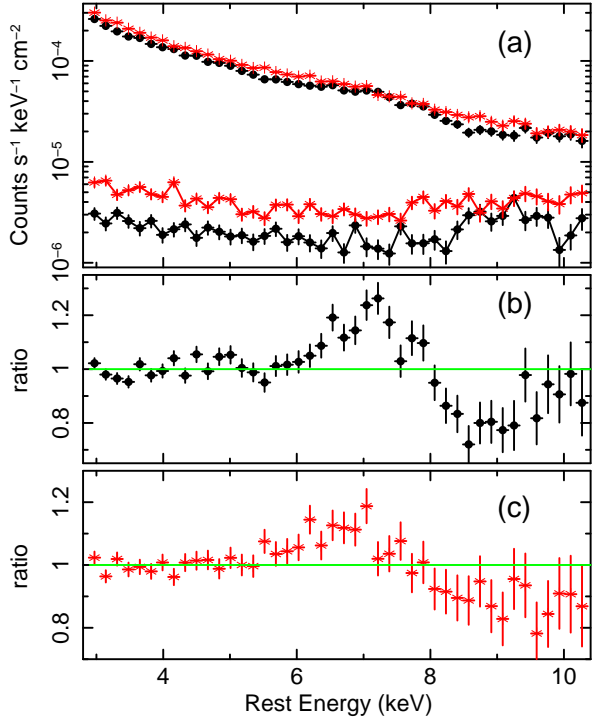}}
\end{center}
\caption{The 2015 {\it XMM-Newton} EPIC-pn spectra of \zw, the OBS\,1 sequence is shown in black, the OBS\,2 sequence in red. The plot is the QSO rest frame at $z=0.0611$. The upper panel (a) shows the net count rate spectra for both observations, while the lower points show the level of the background which lies well below the spectra.
The overall flux of the OBS\,2 spectrum is about 10\% higher than OBS\,1.
The lower panels show data/model ratio of these spectra to a simple power-law model of photon index $\Gamma=2.13\pm0.03$, whereby the OBS\,1 spectrum is shown in panel (b) and the OBS\,2 spectrum is shown in panel (c). A broad emission component is present in the residuals, peaking at around 7\,keV, likely arising from 
ionized iron. At higher energies, a broad absorption trough is present above 8\,keV. The residuals appear to be stronger in the lower-flux OBS\,1 spectrum and shallower in OBS\,2.  }
\label{fig:spectra}
\end{figure}

The OBS\,1 and OBS\,2 spectra were fitted simultaneously with a simple power-law continuum, allowing its normalization to vary between the two observations to account for the variation in overall flux, but linking the photon index between them, with $\Gamma=2.13\pm0.03$. A neutral Galactic column of $N_{\rm H}=6.0\times10^{20}$\,cm$^{-2}$ \citep{Kalberla05} was also included. The fit to this simple model is very poor with $\chi_{\nu}^{2}=263.7/87$, rejected with a null hypothesis probability of $P_{\rm N}=1.1\times10^{-19}$. Panels (b) and (c) show the data/model ratio residuals to this power-law, which show strong residuals in the Fe K band. A broadened emission line is present near to 7 keV in both datasets, while at higher energies above 8\,keV a broad absorption trough is also present. Comparison between the residuals of the two spectra shows that the line residuals appear to be somewhat stronger in the lower flux OBS\,1 spectrum and somewhat shallower (or broader) in OBS\,2.

\subsection{Gaussian Fe K profile}

To provide an initial parameterization of the Fe K profile, a double Gaussian profile was fitted to the datasets, to account for both the excess emission as well as the absorption trough, where the latter is allowed to have a negative normalization. The line widths of the emission vs absorption components were assumed to be the same for simplicity, although the overall width was allowed to vary between the OBS\,1 and OBS\,2 spectra to account for any profile variability. The line normalizations were also allowed to vary, however the centroid energies of the emission and absorption components were tied between the OBS\,1 and OBS\,2 spectra to reduce any parameter degeneracy. 

\begin{deluxetable}{lcc}
\tablecaption{Iron K profile parameters for \zw.}
\tablewidth{250pt}
\tablehead{
& \colhead{OBS\,1} & \colhead{OBS\,2}}
\startdata
Gaussian emission:-\\
$E_{\rm rest}$$^{a}$ & $7.05^{+0.21}_{-0.19}$ & $7.05^t$ \\
$\sigma^{b}$ & $0.66^{+0.17}_{-0.16}$ & $0.94^{+0.19}_{-0.17}$\\
Normalization$^c$  & $1.85^{+0.78}_{-0.54}$ &  $1.31^{+0.75}_{-0.47}$ \\
Line Flux$^{d}$ & $2.1^{+0.9}_{-0.6}$ &  $1.5^{+0.8}_{-0.5}$ \\
EW$^{e}$ & $390^{+160}_{-110}$ &  $240^{+140}_{-90}$ \\
$\Delta\chi^{2}$$^f$ & 119.3 & -- \\
\hline
Gaussian absorption:-\\
$E_{\rm rest}$$^{a}$ & $8.66^{+0.20}_{-0.22}$ & $8.66^t$ \\
$\sigma^{b}$ & $0.66^t$ & $0.94^t$\\
Normalization$^c$  & $-1.28^{+0.37}_{-0.45}$ &  $-1.05^{+0.42}_{-0.50}$ \\
Line Flux$^{d}$ & $-1.8^{+0.5}_{-0.6}$ &  $-1.5^{+0.6}_{-0.7}$ \\
EW$^{e}$ & $-410^{+120}_{-140}$ &  $-290^{+120}_{-140}$ \\
$\Delta\chi^{2}$$^f$ & 62.2 & --\\
\hline
Continuum:-\\
$\Gamma$ & $2.14\pm0.04$ & $2.14^{t}$\\ 
$F_{\rm 2-10\,keV}$$^{g}$ & 5.22 & 6.04\\
\hline
$\chi_{\nu}^{2}$$^{h}$ & $81.8/81$\\
\hline
\hline
P-cygni: \\
$E_{0}$$^{a}$ &$7.3\pm0.1$& $7.1^{+0.1}_{-0.2}$ \\
$\alpha_1$ &$1.9^{+1.6}_{-1.1}$&$1.9^{t}$\\
$v_{\infty}/c$&$-0.35 ^{+0.03}_{-0.04}$& $<-0.39$\\
$\tau_{\rm tot}$&$0.19^{+0.05}_{-0.04}$&$0.11^{+0.03}_{-0.03}$\\
$N_{\rm H}$$^{i}$ & $6.3\pm1.6$ & $3.8\pm1.3$\\
\hline
$\chi^{2}/\nu$$^h$ & $78.5/80$
\enddata
\tablenotetext{a}{Rest frame centroid energy in keV.}
\tablenotetext{b}{Gaussian width in keV.}
\tablenotetext{c}{Gaussian normalization (photon flux) in units $\times10^{-5}$\,photons\,cm$^{-2}$\,s$^{-1}$.}
\tablenotetext{d}{Line flux in units $\times10^{-13}$\,ergs\,cm\,s$^{-1}$.}
\tablenotetext{e}{Equivalent width in eV.}
\tablenotetext{f}{Improvement in $\chi^2$ upon adding component to model.}
\tablenotetext{g}{Observed 2--10\,keV flux in units of $\times10^{-12}$\,erg\,cm$^{-2}$\,s$^{-1}$.}
\tablenotetext{h}{Reduced chi-squared for fit.}
\tablenotetext{i}{Hydrogen column density in units of $\times10^{-23}$\,cm$^{-2}$, calculated from the profile optical depth ($\tau_{\rm tot}$).}
\tablenotetext{t}{Denotes parameter is tied in fit.}
\label{tab:profile}
\end{deluxetable}

The fit parameters to the Gaussian model are tabulated in Table~2. The addition of both lines significantly improved the fit by 
$\Delta\chi^2=-119.3$ and $\Delta\chi^2=-62.2$ for the emission and absorption respectively, while the overall fit statistic was reduced to an 
acceptable $\chi_{\nu}^2=81.8/81$. The best-fit Gaussian profiles are shown in Figure~2 (panel a), superimposed on the fluxed spectra for OBS\,1 and 
OBS\,2.\footnote{The fluxed spectra in this paper are created using an input count rate spectrum (units of counts\,s$^{-1}$\,keV$^{-1}$\,cm$^{-2}$), which is folded 
by the instrumental response, but has been divided through by the instrumental effective area (using the \textsc{setplot area} command within \textsc{xspec}). The y-axis values are then multiplied twice by energy and by a conversion factor of 
$1 {\rm keV} = 1.602\times10^{-9}$\,ergs to convert the spectrum into $\nu F_{\nu}$ flux units.}
The Gaussian profiles returned rest frame centroid energies of $E=8.66^{+0.20}_{-0.22}$\,keV 
in absorption and $E=7.05^{+0.21}_{-0.29}$\,keV for the emission. The centroid energy of the absorption trough implies it is substantially blue-shifted if it is associated with the strong $1s\rightarrow2p$ lines of highly ionized iron. Compared to the expected lab frame energies of the He-like Fe\,\textsc{xxv} resonance line (at 6.7\,keV) or the H-like Fe\,\textsc{xxvi} Lyman-$\alpha$ line (at 6.97\,keV), then the corresponding outflow velocities are $v/c=-0.24\pm0.02$ and $v/c=-0.21\pm0.02$ respectively. On the other hand, the centroid of the emission is consistent with an origin from H-like iron. The profile is also broadened, 
with a best-fit Gaussian width of $\sigma=660^{+170}_{-160}$\,eV for OBS\,1 versus $\sigma=940^{+190}_{-170}$\,eV for OBS\,2. However, given the errors, the difference in velocity width between the profiles is only significant at about the 90\% confidence level and both profiles can also be adequately fitted with a common velocity width, of $\sigma=770\pm160$\,eV. Note that this line width, with respect to the emission centroid at 7\,keV, corresponds to a velocity broadening of $\sigma_{\rm v}=33\,000$\,km\,s$^{-1}$ (or $0.11c$). Overall, the profile appears very reminiscent of the broad, P Cygni like profile measured from the fast wind in PDS\,456 (\citealt{Nardini15}).

Notably the equivalent widths of the emission vs absorption components are roughly equal; for instance during OBS\,1 the equivalent width of the emission line ($390^{+160}_{-110}$\,eV) is similar to the absorption trough ($-410^{+120}_{-140}$\,eV), while in OBS\,2 the equivalent widths are slightly smaller (see Table\,2).  
If the emission originates via re-emission from a wind, this implies that the geometrical covering of the wind is relatively high, as most of the continuum photons that are absorbed by material covering a substantial fraction of $4\pi$ steradians are subsequently re-emitted. On the other hand, if the absorbing gas was isolated to a relatively small clump of material located only along the line of sight, then its total emission would be relatively small and the Fe K profile would be narrow. We explore this further below, where we model the Fe K profile with a P Cygni profile from a near spherical wind. 

\begin{figure}
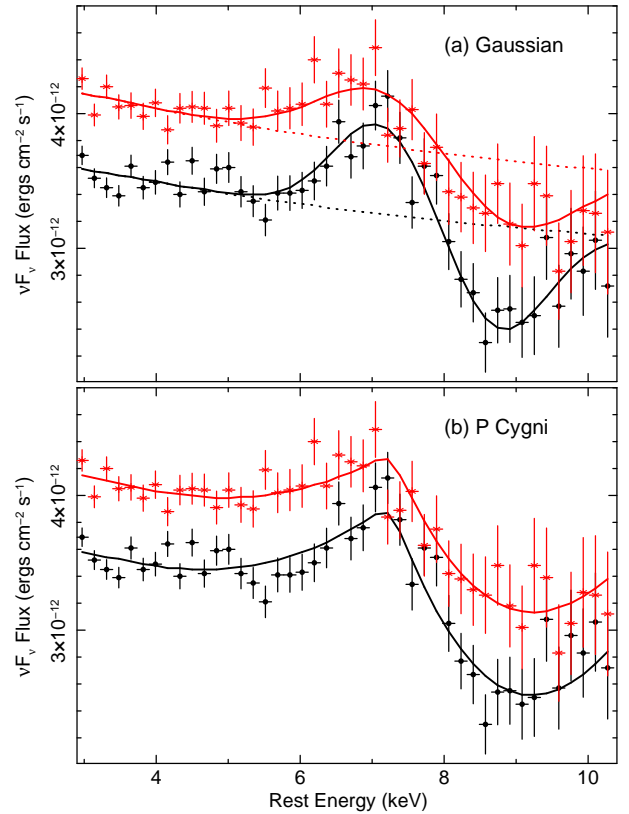

\begin{center}
\rotatebox{-90}{\includegraphics[height=8.5cm]{fig2a.ps}}
\rotatebox{-90}{\includegraphics[height=8.5cm]{fig2b.ps}}
\end{center}
\caption{Fluxed spectra of \zw\ fitted with two profiles; (a) a Gaussian profile in emission and absorption and (b) a P Cygni wind profile. OBS\,1 spectra are in black and OBS\,2 spectra in red. The Gaussian profile in panel (a) clearly shows the broadened emission vs blueshifted absorption in the iron K band, while the best-fitting power-law continuum is shown as a dashed line. The best-fit centroid energy of the blueshifted absorption trough, near to 8.6\,keV, when compared to the expected energies of He or H-like iron (6.70 or 6.97\,keV), indicate the presence of a fast wind.
The P Cygni profile in panel (b) is fitted with the model of \citet{Done07}  from a spherical wind reaching a terminal velocity of 
$v_{\infty}\sim0.35c$. See Section 3 and Table 2 for details of the model parameters.}
\label{fig:profile}
\end{figure}

\subsubsection{Comparison with EPIC-MOS}

The spectra obtained from the EPIC-MOS cameras were also checked for consistency with the EPIC-pn. After the individual MOS\,1 and MOS\,2 spectra were found to be consistent, these were combined into a single MOS spectrum for each observation, after combining the response files with the appropriate weighting. Figure~3 shows the resulting MOS spectrum shown for OBS\,1 versus the pn spectrum, where the upper-panel shows a ratio to a simple power-law model.  
Both spectra are clearly consistent with each other, with the MOS data also showing both the broad emission component centered near 7 keV and a broad blue-shifted absorption trough near to 9\,keV. A joint fit between the pn and MOS spectra for OBS\,1 yielded consistent result compared to above, where for the absorption line $E=8.7\pm0.2$\,keV, $\sigma=635\pm150$\,eV and ${\rm EW} = -360\pm85$\,eV, with consistent parameters also obtained for the broad ionized emission. No further residuals are present in either the pn or MOS spectra at iron K, once these Gaussian components are included in the model (see lower panel, Figure~3).
Likewise the MOS spectra for OBS\,2 are also consistent with the pn. Thus both the pn and MOS data verify the presence of the blue-shifted absorption in \zw.

\begin{figure}
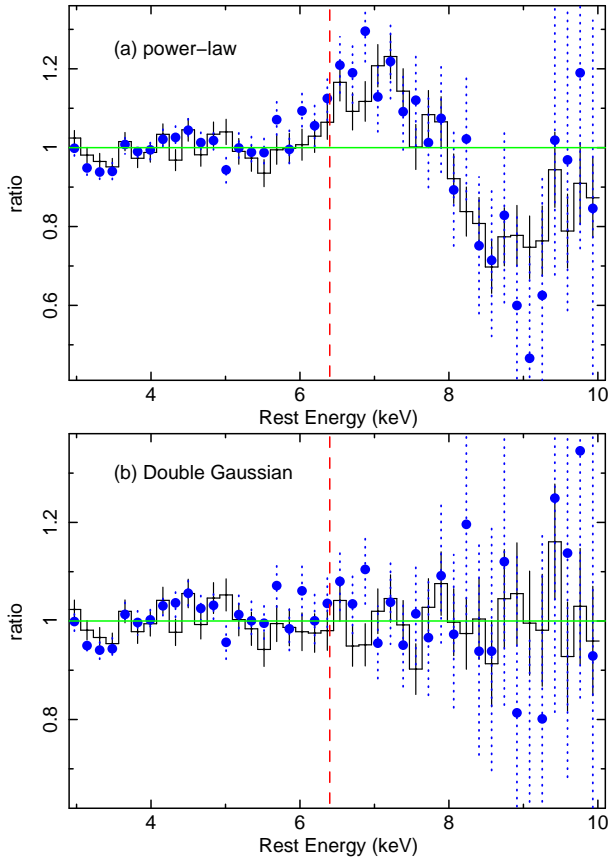

\begin{center}
\rotatebox{-90}{\includegraphics[height=8.5cm]{fig3.ps}}
\rotatebox{-90}{\includegraphics[height=8.5cm]{fig3b.ps}}
\end{center}
\caption{Comparison between the pn and MOS spectra for OBS\,1, where the pn data are in black and the MOS data are shown as blue circles.
The upper panel shows the ratio of the spectra compared to a power-law with Galactic absorption. The broad Fe K emission line and blue-shifted absorption feature, centered near 7 keV and 9\,keV respectively, are detected independently in the MOS spectra and the overall profile is consistent between the pn and MOS. Note that both datasets are binned to constant energy bins of $\Delta E = 160$\,eV, at approximately the FWHM resolution. The lower panel shows the residuals after the broad Gaussian absorption and emission components have been added to the model. The dashed red-line marks the expected position of a neutral Fe K$\alpha$ line, which is not observed in the spectra.}
\label{fig:MOS}
\end{figure}

We also attempted to place limits on any narrow Fe K$\alpha$ emission in the \zw\ spectra. Narrow components of the 6.4\,keV Fe K$\alpha$ fluorescence line appear to be almost ubiquitous in the X-ray spectra of Seyfert galaxies (e.g. \citealt{Nandra07}) and may originate from X-ray reflection off distant Compton thick matter, such as a pc scale molecular torus (\citealt{MurphyYaqoob09,Ikeda09,BrightmanNandra11}). 
The simultaneous pn and MOS spectra for each observation were used to place a limit on the narrow iron K$\alpha$ component at 6.4\,keV. 
As can be seen in Figure\,3 (lower panel), once the broad ionized emission and absorption lines are included in the model, there are no residuals apparent near to 6.4\,keV in either the pn or MOS spectra. 
To calculate an upper limit on the equivalent width, a narrow Gaussian was included with a fixed width of $\sigma=10$\,eV, while the centroid energy of the Gaussian was restricted to within $\pm0.1$\,keV of 6.4\,keV.  A best-fit continuum model consisting of a power-law and the two broad Gaussians was adopted, allowing the fit parameters to adjust accordingly. 
A tight upper-limit on the equivalent width of ${\rm EW}<30$\,eV was found for OBS\,1, while for OBS\,2 the upper-limit is lower still, with ${\rm EW}<18$\,eV.  
Thus as the contribution of a distant reflection component, via the neutral Fe K$\alpha$ line, appears negligible in \zw, it has not been included in any of the subsequent modeling.

The weakness of the narrow Fe K$\alpha$ line in \zw\ might be explained by the X-ray Baldwin effect, where the equivalent width of the narrow Fe K$\alpha$ line is observed to decrease with increasing AGN X-ray luminosity (\citealt{IwasawaTaniguchi93,Nandra97,ReevesTurner00,Page05,Bianchi07}). Note that for the 2-10\,keV luminosity of \zw\ of $L_{\rm 2-10 keV}=5\times10^{43}$\,erg\,s$^{-1}$, \citet{Bianchi07} predicted an equivalent width in the range of $40-90$\,eV (see their Figure 1) 
and thus the above upper limits are somewhat lower than expected. For the likely black hole mass and 
bolometric luminosity of \zw, then its Eddington ratio is of the order $L_{\rm bol}/L_{\rm Edd}\sim 1$. From the anti-correlation in \citet{Bianchi07} between the Fe K$\alpha$ equivalent width and Eddington ratio, the predicted equivalent width is $20-70$\,eV which is consistent with the limits observed. Thus the high Eddington ratio of \zw\ might explain the weakness of its narrow Fe K$\alpha$ line. 

\subsection{P Cygni profile}
Since the observed profiles are reminiscent of the classical P-Cygni like profile, we tested the  same   customized model   for a P-Cygni profile  that was applied by \citet{Nardini15}   to PDS\,456. The model was developed   for the   Fe-K absorption feature seen in 1H\,0707-495  (\citealt
{Done07}) and  is based  on the Sobolev approximation with exact integration (SEI) for  a spherically symmetric wind.   The parameters of the model are: the energy of the onset of the absorption component  ($E_0$), the terminal velocity of the wind ($v_{\infty}$),  how the velocity scales with distance ($\gamma$), the initial velocity at the photosphere ($w_{0}$), the optical depth ($\tau_{\rm tot}$) and the  smoothness of the profile. The smoothness is defined by two parameters $\alpha_1$ and $\alpha_2$, where higher values correspond  to smoother profiles. The velocity field  is defined   by $w=w_0+(1-w_0)(1-1/x)^\gamma$; where  $w=v/v_{\infty}$ is the ratio between the wind velocity and the terminal velocity $v_{\infty}$,  $w_{0}$  is the initial velocity at the photosphere and $x=r/R_{0}$ is the radial distance in units of the   photospheric radius. 
At large radii, where $r>>R_0$, then the wind velocity tends to $v= v_{\infty}$
Following \citet{Nardini15}  we adopted  $\gamma=1$ and  $w_{0}=0.01$,  as they have a marginal effect on the profile, with these parameters  the line optical depth varies as:-
\begin{equation}
 \tau(w)\propto \tau_{\rm tot}\, w^{\alpha_1}\,(1-w)^{\alpha_2}.
\end{equation}

We then replaced the two Gaussians with the P-Cygni model and  fitted   simultaneously OBS\,1 and OBS\,2.  The pn spectra were used as these yield the highest S/N at high energies, but noting the consistency with the MOS. We tied the  underlying continuum slope but allowed the normalizations to vary. Regarding the P-Cygni parameters, we allowed the optical depths, the terminal velocities and the rest frame energies of the P-Cygni line, $E_0$, to vary independently, in order to account for changes in the profiles and their intensities. We assumed  that $\alpha_1=\alpha_2=\alpha$, as they cannot be constrained separately. This results in a symmetrical absorption profile, where the trough minimum (at maximum $\tau$) occurs at $v_{\infty}/2$.
As expected from the earlier Gaussian profile, the  P-Cygni model results in a good fit ($\chi^2/\nu=78.5/80$). The best fit parameters are reported in Table~2. Note that similar to the Gaussian model, the P-Cygni model results in a  marginally shallower profile during OBS\,2, where 
we found that the optical depth marginally decreases from $\tau_{\rm tot}=0.19_{-0.04}^{+0.05}$ (OBS\,1) to  $\tau_{\rm tot}=0.11_{-0.03}^{+0.03}$ (OBS\,2). 
We can only place a lower limit to the terminal velocity ($v_{\rm {\infty}}/c<-0.39$) during OBS\,2, while during OBS\,1 we derive a terminal velocity of 
$v_{\rm {\infty}}/c=-0.35_{-0.04}^{+0.03}$. 

From the optical depth ($\tau_{\rm tot} $) we can derive an estimate of the ionic column density of the gas through the relation $\tau_{\rm tot}=(\pi e^2/m_e\,c)\,f \lambda_0 N_{\rm i}/v_{\infty}$, where $m_e$ and $e$ are the electron mass and charge (in Statcoulomb), $ \lambda_0$ (in cm) is the wavelength of the line (lab frame) and  $f$ is the oscillator strength of the transition. Now, assuming an identification with {Fe\,\textsc{xxvi}}
Ly$\alpha$ ($f=0.21$), we derive ionic columns of $N_{\rm i}=(2.0\pm 0.5)\times 10^{19}$\,cm$^{-2}$  and $N_{\rm i}=(1.2\pm 0.4)\times 10^{19}$\,cm$^{-2}$  for OBS\,1 and OBS\,2, respectively. Thus for the Solar abundances of \citet{GrevesseSauval98}, these correspond to $N_{\rm H}=(6.3\pm1.6)\times 10^{23}$\,cm$^{-2}$   and  $N_{\rm H}=(3.8\pm1.3)\times 10^{23}$\,cm$^{-2}$ for OBS\,1 and  OBS\,2.
 
\section{Photoionization Modelling} \label{sec:photoionization_modelling}

\subsection{The Photo-ionizing Continuum}

We next modeled the \xmm\ spectrum with a self consistent photoionization model,
using the \textsc{xstar} code (\citealt{Kallman04}). 
The UV to X-ray SED of \zw\ was used to estimate the input photo-ionizing continuum, which is plotted in Figure~4 for the 2015 epoch (for simplicity we only used the 
OBS\,1 sequence). 
Simultaneous UV photometry from the UVW1 and UVW2 filters from the Optical Monitor (OM) on-board \xmm\, along with the 
0.3--10\,keV EPIC-pn spectrum, was used. In addition, we adopted an earlier non-simultaneous datapoint from {\it FUSE} in the far UV at $\sim1000$\,\AA\ (\citealt{Scott04}), in order to anchor the UV continuum, noting that the FUSE datapoint lies on the extrapolation of the UV continuum in Figure~4. 
All data-points have been corrected for Galactic extinction of $E(B-V)=0.057$, while the X-ray data points are corrected for Galactic photoelectric absorption, 
corresponding to a column of $N_{\rm H}=6\times10^{20}$\,cm$^{-2}$ and Solar abundances of \citet{GrevesseSauval98}.

\begin{figure}
\begin{center}
\rotatebox{-90}{\includegraphics[height=8.5cm]{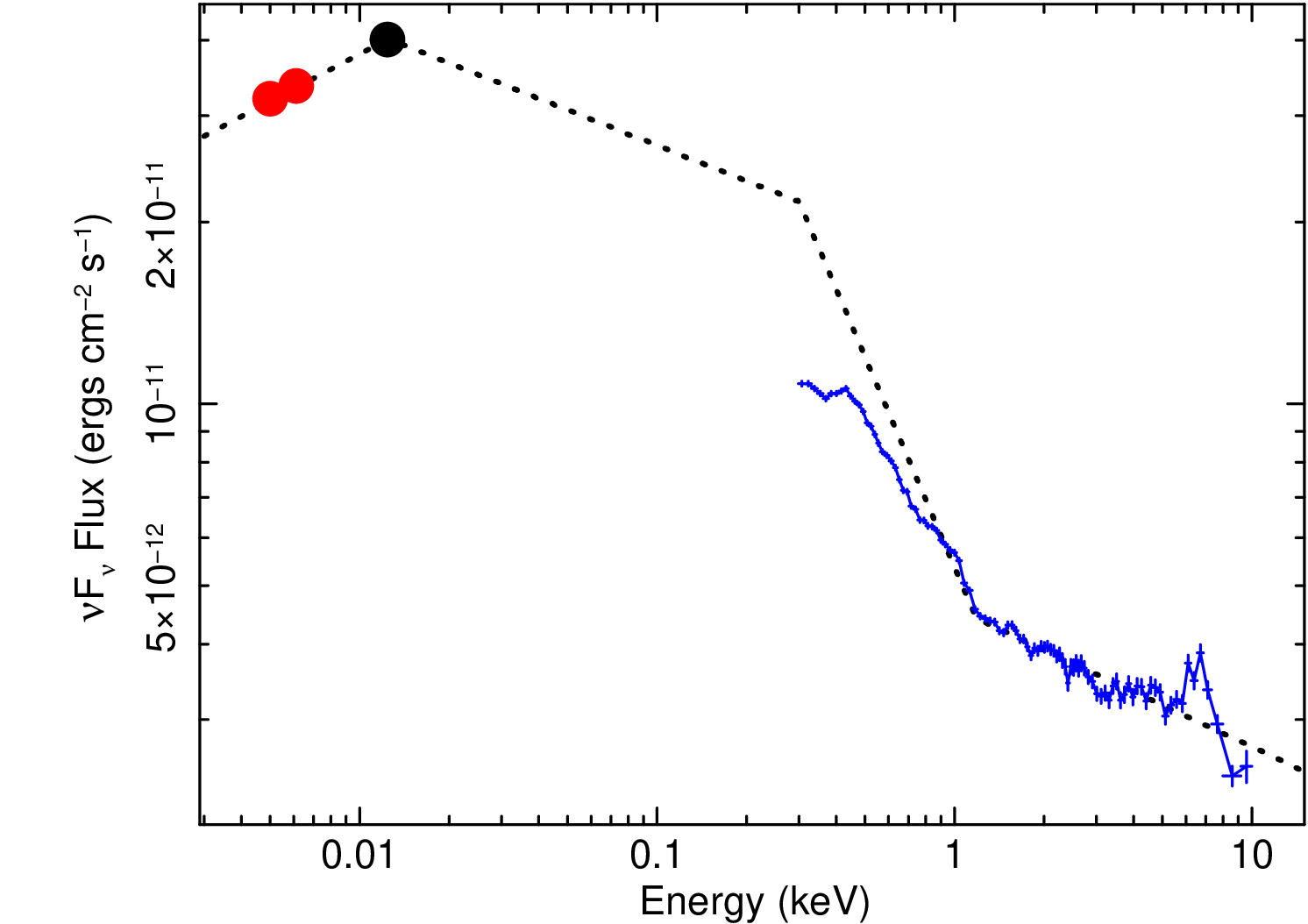}}
\end{center}
\caption{The UV to X-ray SED of \zw, taken from the 2015 \xmm\ observations. The EPIC-pn spectrum (from OBS\,1) is shown in blue, while UV photometric points from the UVW1 and UVW2 filters from the \xmm\ OM are shown as red circles. Note that the black circle shows the non-simultaneous measurement from FUSE in the far UV (\citealt{Scott04}). The UV and X-ray data are corrected for reddening and Galactic absorption respectively. The dashed line shows the phenomenological model fitted to the SED and corrected for intrinsic X-ray absorption. This consists of a series of power-law portions, with 3 break points, at 12.5\,eV, 300\,eV and 1.2\,keV; see Section~4.1 for details.}
\label{fig:sed}
\end{figure}

The SED is parameterized by a series of power-laws, with 3 break-points. Below 12.5\,eV in the UV, the OM and {\it FUSE} points are connected by a photon index of 
$\Gamma_{UV}=1.75$, while the far UV to soft X-ray (from 12.5\,eV to 300\,eV) bands are connected by $\Gamma_{UVX}=2.2$. From 0.3--1.2\,keV, the soft X-ray spectrum is modeled by a steep photon index of $\Gamma_{SX}=3.3\pm0.2$ to approximate the soft excess, while a photon index of $\Gamma_{X}=2.21\pm0.03$ describes the hard X-ray power-law above a break energy of 1.2\,keV. Note that the opacity due to the warm absorber, as modeled by a two phase model in \citet{Silva18}, has been accounted for in determining the soft X-ray continuum. From this SED, the subsequent ionizing ($1-1000$\,Ryd) luminosity is estimated to be $L_{\rm ion}\sim1.5\times10^{45}$\,erg\,s$^{-1}$. This is likely to be a relatively conservative estimate for the ionizing luminosity, which could be somewhat higher, e.g. if the SED peaks in-between the observable UV and soft X-ray bands. 
If we adopted a higher break energy, of 100\,eV (instead of 12.5\,eV) for the first breakpoint between the UV and soft X-rays, then the 1-1000\,Ryd band luminosity is slightly higher ($3\times10^{45}$\,erg\,s$^{-1}$). 
We note that \citet{Porquet04} also estimated a total bolometric luminosity of $3\times10^{45}$\,erg\,s$^{-1}$, based on scaling the 5100\,\AA\ flux, which is close to the Eddington value for the black hole mass of \zw. 

Regardless of the exact parameterization of the overall SED, the most critical parameter for the photoionization modeling 
is the X-ray photon index above 1\,keV, which sets the ionization balance for the highly ionized iron K-shell lines. 
Thus if the X-ray continuum was much harder ($\Gamma<2$) than observed here, then the number of ionizing photons above the Fe K edge threshold at 7.11\,keV would be greater and the Fe K features will be subsequently more ionized and weaker, as more ions become fully ionized. Future observations with {\it NuSTAR} will be able to determine the exact form and slope of the continuum above 10\,keV, although the pn spectra suggest a steep hard X-ray slope with $\Gamma=2.1-2.2$. 

\begin{deluxetable*}{lcccc}
\tablecaption{Photoionization Modeling of the Wind.}
\tablewidth{500pt}
\tablehead{\colhead{Parameter} & \colhead{2015 OBS\,1} & \colhead{2015 OBS\,2} & \colhead{2002} & \colhead{2005}}
\startdata
Fe K absorber:-\\
$N_{\rm H}$$^{a}$ & $7.5^{+1.4}_{-1.2}$ & $4.5^{+1.4}_{-1.1}$ & $4.8^{+2.3}_{-2.0}$ & $2.0^{+1.2}_{-1.1}$\\
$\log\xi$$^{b}$ & $4.91^{+0.37}_{-0.13}$ & $4.91^t$ & $4.9^{f}$ & $5.0^t$\\
$v/c$  & $-0.265\pm0.010$ & $-0.265^t$ & $-0.29\pm0.03$  & $-0.25\pm0.02$\\
$F_{\rm abs}$$^c$ & $4.0^{+0.7}_{-0.6}$ & $2.9^{+0.9}_{-0.7}$ & $4.7^{+2.0}_{-1.7}$ & $<0.95$ \\
\hline
Fe K Emission:-\\
$N_{\rm H}$$^{a}$ & $7.5^t$ & $4.5^t$ & $4.8^t$ & $4.6^{+1.7}_{-1.3}$\\
$\log\xi$$^{b}$ & $4.91^t$ & $4.91^f$ & $4.9^f$ & $5.0^{+0.5}_{-0.4}$ \\
$F_{\rm emiss}$$^c$ & $1.9^{+0.6}_{-0.4}$ & $1.3^{+0.6}_{-0.5}$ & $<4.1$ & $1.8^{+0.7}_{-0.5}$\\
$\kappa^{d}$ & $3.7^{+1.2}_{-0.9}\times10^{-4}$ & $4.2^{+2.2}_{-1.8}\times10^{-4}$ & $<8.2\times10^{-4}$ & $5\times10^{-4}$$^f$\\
$f=\Omega/4\pi^{d}$ & $0.71^{+0.23}_{-0.17}$ & $0.81^{+0.19}_{-0.35}$ & -- & $1.0^f$\\
\hline
Continuum:-\\
$\Gamma$ & $2.15\pm0.03$ & $2.15^f$ & $2.22\pm0.07$ & $2.10\pm0.05$ \\
$F_{\rm 2-10\,keV}$$^{e}$ & 5.24 & 6.08 & 8.47 & 5.01
\enddata
\tablenotetext{a}{Units of column density $\times10^{23}$\,cm$^{-2}$. }
\tablenotetext{b}{Ionization parameter (where $\xi=L/nR^{2}$) in units of erg\,cm\,s$^{-1}$.}
\tablenotetext{c}{Flux absorbed from the continuum or re-emitted over the Fe K band, in units $\times10^{-13}$\,erg\,cm$^{-2}$\,s$^{-1}$.}
\tablenotetext{d}{Fitted normalization of \textsc{xstar} emission component, where $\kappa= f L_{38}/D_{\rm kpc}^2$ and $f=\Omega/4\pi$ is the emitter covering fraction, $L_{38}$ is the 1--1000\,Ryd ionizing luminosity in units of $10^{38}$\,erg\,s$^{-1}$ and $D_{\rm kpc}$ is the distance to \zw\ in units of kpc.}
\tablenotetext{e}{Observed 2--10\,keV flux, not corrected for absorption, in units of $\times10^{-12}$\,erg\,cm$^{-2}$\,s$^{-1}$.}
\tablenotetext{f}{Denotes parameter is fixed.}
\tablenotetext{t}{Denotes parameter is tied between observations.}
\label{tab:xstar}
\end{deluxetable*}

\subsection{Photoionization Results}

Grids of photoionization models were subsequently generated within \textsc{xstar} for the spectral fitting, using the above SED as the input continuum. 
The absorption was accounted for by a multiplicative grid, while the emission from the wind was modeled by an additive grid. 
A velocity broadening of $b=25\,000$\,km\,s$^{-1}$ was used in the models, accounted for by the turbulence velocity parameter\footnote{Within \textsc{xstar}, the turbulence velocity is defined as $b=\sqrt{2} \sigma = {\rm FWHM}/(2\sqrt{\ln 2}$).} and is consistent with the line widths inferred from the earlier Gaussian analysis.
Solar abundances of \citet{GrevesseSauval98} were used throughout. The overall form of the model is:-

\begin{equation}
{\rm tbabs} \times ({\rm xstar}_{\rm abs} \times {\rm pow} + {\rm xstar}_{\rm emiss})
\end{equation}

\noindent where ${\rm xstar}_{\rm abs}$ denotes the iron K absorption and ${\rm xstar}_{\rm emiss}$ represents the 
photoionized emission.
The spectra are absorbed by a Galactic component of absorption, via the \textsc{tbabs} model (\citealt{Wilms00}), as above.
Note that we allowed the column density to vary between the 
OBS\,1 and OBS\,2, spectra, but tied the ionization and outflow velocity between them, which otherwise were consistent within errors. We assumed that the 
column density of the emission component (as well as its ionization), to be the same as the absorber and these were subsequently tied. Note that the net outflow velocity of the emitter was not tied to that the absorber. In this case no strong net blue-shift was required for the emitter, with an upper-limit of $v<0.05c$ and thus was subsequently fixed at zero. We note that in the case of a wide angle wind, the observed emission can be observed over all angles and thus a net blue-shift of the emission need not be observed. In this respect, both the P Cygni model and the disk wind model of \citet{Sim08,Sim10a} investigated later (see Section 5) self-consistently calculate the expected velocity profiles for a spherical and bi-conical wind geometry respectively. 

The best-fit parameters of the photoionized emission and absorption model are shown in Table~3 and overall the fit statistic is good, with $\chi_{\nu}^2=71.8/81$. 
Removing either the absorption or emission from the model results in a substantially worse fit, with $\Delta\chi^2=91$ for $\Delta\nu=3$ for the absorber versus 
$\Delta\chi^2=101$ for $\Delta\nu=3$ for the emission. 
Figure~5 shows the best fit \textsc{xstar} model fitted to both spectra, which can well reproduce both the emission and absorption from the wind. 
The column density from the OBS\,1 spectrum ($N_{\rm H}=7.5^{+1.4}_{-1.2}\times10^{23}$\,cm$^{-2}$) was found to be slightly higher than for OBS\,2 ($N_{\rm H}=4.5^{+1.4}_{-1.1}\times10^{23}$\,cm$^{-2}$), which is consistent with the absorption trough being deeper in OBS\,1. These column densities are also consistent with what was 
derived from the earlier P Cygni profile results.
The ionization of the gas is high, with $\log\xi=4.91^{+0.37}_{-0.13}$, with the absorption and emission 
mainly arising from He and H-like iron. The outflow velocity of the absorber was $-0.265\pm0.010c$, while no significant velocity shift was required in emission. Note that the outflow velocity derived from the \textsc{xstar} model is slightly lower than for the terminal velocity obtained from the P Cygni model, as it is calculated from the centroid of the absorption profile, rather than from its maximum bluewards extent.

\begin{figure}
\begin{center}
\rotatebox{-90}{\includegraphics[height=8.5cm]{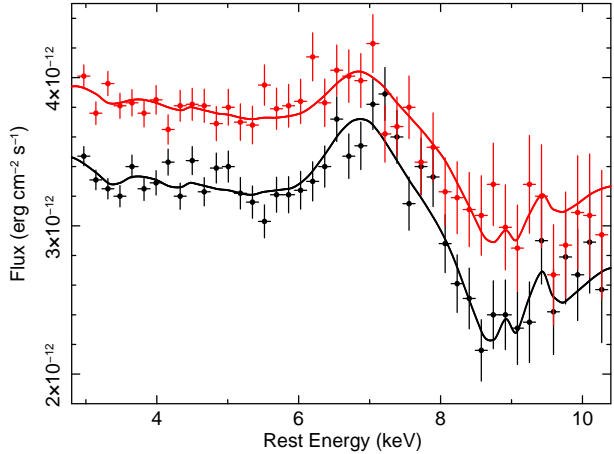}}
\end{center}
\caption{The 2015 pn spectra of \zw, fitted with the \textsc{xstar} model described in Section\,4.2 and with the parameters listed in Table\,3.
The emission and absorption model applied to the data has a velocity broadening of 25\,000\,km\,s$^{-1}$, in order to account for 
the profile width.
A fast absorber, of outflow velocity $v/c=-0.265\pm0.010$, accounts for the broad absorption trough, while the iron K emission results from the subsequent 
re-emission from the wind. Note the very weak emission feature at 9.3\,keV is the radiative recombination emission from Fe\,\textsc{xxvi}.}
\label{fig:xstar}
\end{figure}

The geometrical covering fraction of the gas $f$ was also calculated from the normalization of the \textsc{xstar} emission component $\kappa$, where:-
\begin{equation}
\kappa = f \times \frac{L_{38}}{D_{\rm kpc}^2}.
\end{equation}
Here, $L_{38}$ is the 1--1000\,Rydberg ionizing luminosity and $D_{\rm kpc}$ is the luminosity distance to the source in kpc, while $f=1$ for a fully covering spherical shell of gas.
Thus from the measured normalizations reported in Table\,3 and adopting $D=240$\,Mpc and $L=3\times10^{45}$\,erg\,s$^{-1}$ for \zw, the 
covering fraction was estimated to be $f=0.71^{+0.23}_{-0.17}$ and $f=0.81^{+0.19}_{-0.35}$ for OBS\,1 and OBS\,2 respectively. 
The gas covering can also be estimated by comparing the flux absorbed from the continuum with what is re-emitted over the iron K band in the 
form of line emission, where a ratio of close to one might be expected for a fully covering wind. These values are reported in Table\,3, which shows that about half of the incident radiation that is absorbed is subsequently re-emitted. Thus the wind is likely to cover at least $2\pi$ steradians solid angle with respect to the X-ray continuum source. 

\subsection{Alternative Reflection Models}

We also tested whether the Fe K profile of \zw\ could be fitted with relativistically blurred reflection instead of a wind, which then could originate off the surface of the inner accretion disk. 
Initially a model with no wind absorption was tested, simply consisting of an ionized blurred reflection component and a power-law continuum, which are absorbed only by the 
Galactic absorption.
To model the reflection, the \textsc{xillver} emission table was used \citep{Garcia13}, which was convolved with relativistic blurring kernel, \textsc{kdblur}, 
which approximates the disk emissivity versus radius with a simple power-law function as $R^{-q}$. The inner disk radius was initially fixed to the innermost radius expected from around a maximal Kerr black hole (with $R_{\rm in}=1.24 R_{\rm g}$), while the outer radius was fixed to  $R_{\rm out}=400 R_{\rm g}$. The iron abundance of the reflector was assumed to be Solar, 
which is otherwise poorly constrained with an upper-limit of $A_{\rm Fe}<5$. The continuum incident upon the reflector was assumed to have the same photon index as the primary power-law, while the high energy cut-off was fixed at 300\,keV, as this cannot be constrained without hard X-ray data. The normalization and ionization of the reflector was allowed to vary, as well as the disk inclination and emissivity index, while the power-law photon index and normalization were allowed to vary for both observations.
However, this reflection model resulted in a rather poor fit, with $\chi_{\nu}^2=125.2/79$ and the model left significant residuals above 8\,keV due to the presence of the blue-shifted absorption.

Thus the input model was adjusted to include the photoionized absorption from the wind, as well as the reflected emission from the accretion disk. The 
form of the model is then:-

\begin{equation}
{\rm tbabs} \times ({\rm xstar}_{\rm abs} \times {\rm pow} + {\rm kdblur} \otimes {\rm xillver}).
\end{equation}

\noindent Note that this is phenomenologically identical to the above \textsc{xstar} emission plus absorption model, except that the photoionized emission has been replaced by the 
ionized reflector (\textsc{xillver}), which is convolved with \textsc{kdblur} to account for the line broadening. This model then did provide an acceptable fit, where $\chi_{\nu}^2=73.3/76$ and it appears identical to that in Figure~5, 
The wind parameters remained unchanged within errors from before; e.g. for OBS1\, for an absorber ionization of $\log\xi=5$, the column density is $N_{\rm H}=6.8^{+4.2}_{-2.3}\times10^{23}$\,cm$^{-2}$, with an outflow velocity of $v=-0.24\pm0.01c$. The ionization of the reflector is $\log\xi=3.1\pm0.3$, while the inclination is also quite high, with $\theta=58^{+15}_{-8}\degg$ in order to match the centroid energy of the broad emission component. The emissivity index is relatively flat, with $q=2.2^{+0.6}_{-0.5}$, which suggests that the X-ray emission is not highly centrally concentrated close to the black hole, as is evidenced through the relative lack of a strong red-wing to the iron K profile (see Figure 3). 
Indeed only an upper-limit of $R_{\rm in}<16 R_{\rm g}$ can be placed upon the inner disk radius in this case. 
In this model, the reflection fraction is constrained to $R=0.6^{+0.3}_{-0.2}$\footnote{Here the reflection fraction is defined as the ratio of the reflected flux to that of the incident power-law, calculated over the 3--100\,keV energy range.}, while the photon index is $\Gamma=2.17\pm0.05$. 

Overall, a contribution of a broad disk reflection component to the iron K profile cannot be excluded, although the presence of a fast disk wind is still required in any event to account for the deep 9\,keV absorption trough. In the context of disk reflection, the low emissivity index could be consistent with a more extended X-ray corona and indeed an extended coronal component has been suggested in \zw\ by \citet{Wilkins17} based upon its X-ray timing properties. It should be noted that the wind itself can also produce significant X-ray reflection via scattering off the wind surface. This latter contribution will be investigated later in Section~5, using the physically motivated accretion disk wind models of \citet{Sim08,Sim10a}. These self consistently compute both the line of sight absorption from the wind, as well as the scattered wind emission integrated over all angles and accounting for relativistic effects.

\subsection{Wind Variability}

The earlier \xmm\ datasets of \zw\ were also compared to the above \textsc{xstar} model to place any additional constraints on the possible wind, as well as any long term variability. 
These 2002 and 2005 datasets were originally analysed by \citet{Gallo04}   and \citet{Porquet04}  (for the 2002 observation) and by \citet{Gallo07}   (for the 
2005 observation) and in all of these analyses, a broad ionized iron K emission line was found. 
We re-extracted these EPIC-pn spectra from these observations as per Section~2, yielding net exposures of 18\,ks and 57\,ks for 2002 and 2005 respectively (see Table~1), shorter than the 2015 observations. 
The background level was low in both of these observations. As per the 2015 observations, the spectra were binned into constant energy intervals and with a minimum signal to noise ratio of 5, which resulted in the short 18\,ks spectrum having a coarser binning of $\Delta E = 240$\,eV per bin.

The ratio of these spectra to a simple power-law model (absorbed only by the neutral Galactic absorber) are shown in Figure~6. Although it is of lower signal to noise, the 2002 spectrum displays residuals
similar to the two 2015 observations, with an excess of emission due to ionized iron around 7\,keV and a broad absorption trough centered near to 9\,keV. 
Application of the \textsc{xstar} model in Section~4.2, with a 25\,000\,km\,s$^{-1}$ velocity broadening, yielded a very good fit to the 2002 spectrum. The best-fit column density of $N_{\rm H}=4.8^{+2.3}_{-2.0}\times10^{23}$\,cm$^{-2}$ and outflow velocity of $v/c=-0.29\pm0.03$ were consistent with the 2015 spectra and overall 
the addition of the fast absorber improved the fit statistic by $\Delta \chi^2=17$ for $\Delta\nu=2$. 
The only apparent difference with the 2015 observations is the continuum flux, which was about 50\% higher in 2002 (see Table~3 for details). 
Note that the ionization parameter was fixed at $\log\xi=4.9$, as per the 2015 spectra, as otherwise it was poorly constrained due to the short exposure. 

\begin{figure}
\begin{center}
\rotatebox{-90}{\includegraphics[height=8.5cm]{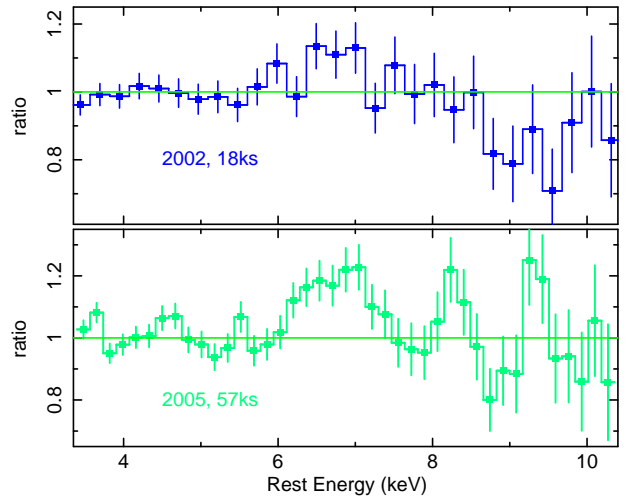}}
\end{center}
\caption{Data/model ratio for the 2002 (top) and 2005 (lower) spectra, against a simple power law model. The residuals in the 2002 spectrum are similar to the 2015 
spectra, with a broad absorption trough near 9\,keV. In contrast, the 2005 spectrum appears dominated by ionized emission, 
both from a broad Fe Ly$\alpha$ emission line near 7\,keV, as well as possible higher order features near 8.3 and 9.3\,keV. }
\label{fig:ratio}
\end{figure}

In contrast, the residuals of the 2005 spectrum to a power-law continuum appear quite different to the other spectra. 
While broad ionized emission is present near to 7\,keV, emission like residuals also appear to be present at higher energies, notably at 8.3\,keV and 9.3\,keV in the 
AGN rest frame, while no strong absorption trough is present. Overall, the fit to a simple power-law model resulted in a very poor fit to the 2005 spectrum, 
with $\chi_{\nu}^2=81.5/41=1.99$, rejected at $>99.99$\% confidence. The 2--10\,keV flux is similar to the 2015 observations.
Most of the contribution towards $\chi^2$ arises from the broad Fe K$\alpha$ component near 7\,keV ($\Delta\chi^2=35$ for $\Delta\nu=3$ when fitted with a Gaussian), while the higher energy features are more marginal (with $\Delta\chi^2=7$ and $\Delta\chi^2=6$ for the 8.3 and 9.3\,keV features respectively). However, despite their low significance, the rest frame energies of the high energy features are entirely consistent with the expected emission from the higher order Fe\,\textsc{xxvi} Ly$\beta$ line and the respective radiative recombination continuum feature from H-like iron.

The 2005 spectrum was first fitted with the 25\,000\,km\,s$^{-1}$ \textsc{xstar} grid, in order to model the emission features. However, the velocity broadening was too large to account for the residuals. As a result, we generated another grid of photoionized spectral models within \textsc{xstar}, but with a 
lower velocity broadening of 10\,000\,km\,s$^{-1}$. Given that this spectrum appears more dominated by the emission, we uncoupled the emitter column from that of the absorber. For the emitter, we fixed the normalization of the \textsc{xstar} component such that it corresponds to full covering with $f=1$, but allowed the 
emitter column to vary. The ionization of the emitter and absorber were tied, as per the above analysis.
Application of this model to the 2005 spectrum resulted in an acceptable fit, with 
$\chi_{\nu}^2=44.7/37$ and was able to account for the Fe K emission features.
The gas ionization is high, with $\log\xi=5.0^{+0.5}_{-0.4}$, consistent with most of the emission arising from H-like iron as noted above, while the 
emitter column was found to be $N_{\rm H}=4.6^{+1.7}_{-1.3}\times10^{23}$\,cm$^{-2}$. The flux of the emission component (see Table~3) is similar to that observed in
2015, which suggests it may be more apparent against the continuum in 2005 which is less absorbed. 
Indeed, the column density of any absorption was found to be about 
a factor of $\times2-3$ lower (with $N_{\rm H}=2.0^{+1.2}_{-1.1}\times10^{23}$\,cm$^{-2}$) compared to 2015. 
Thus overall, three out of the four \zw\ spectra, from the 2002 and 2015 epochs, show evidence for a blue-shifted iron K absorption trough, while the 
2005 spectrum is dominated by the ionized iron K emission. 

\subsubsection{Short Term Variability}

\zw\ shows substantial short-timescale X-ray variability within the longer 2015 observations \citep{Wilkins17} and thus we investigated whether there were any changes in the wind properties in response to the continuum variations.
Due to the mosaic mode used, the two observations were split into $2\times5$ 
sequences, each corresponding to a different telescope pointing. 
While the individual sequences were too short ($\sim15-20$\,ks net exposure) to investigate the wind variability between each pointing, 
we did attempt to extract flux selected spectra from a combination of these pointings. 
To do this, the 10 sequences across OBS\,1 and OBS\,2 were sorted by their 2--10\,keV flux and the three brightest (or faintest) spectra were combined to create a high (or low) flux selected spectrum. For the high flux spectrum, these consisted of the 1st, 4th and 5th observations from OBS\,2, while the low flux spectrum consisted 
of the 3rd, 4th and 5th spectra all from OBS\,1. The net exposures and 2--10\,keV fluxes of the high and low spectra are 52.1\,ks vs 54.3\,ks and $6.4\times10^{-12}$\,erg\,cm$^{-2}$\,s$^{-1}$ vs $4.7\times10^{-12}$\,erg\,cm$^{-2}$\,s$^{-1}$ respectively. 

The high vs low flux spectra are shown in Figure~7. While the spectral shape and profile remain consistent between the spectra, the features appear weaker in the high flux 
spectrum. To quantify these differences, the spectra were modelled with the above \textsc{xstar} model, accounting for any changes by allowing the column density of the 
wind to vary, as well as the normalization of the power-law continuum, while the wind ionization was assumed to remain constant (where $\log\xi=4.90^{+0.12}_{-0.08}$).  
In this case, the column density in the high flux spectrum was found to be significantly lower, with 
$N_{\rm H} = 4.1^{+1.5}_{-1.3}\times10^{23}$\,cm$^{-2}$ compared to the low flux case with $N_{\rm H} = 8.1^{+2.0}_{-1.6}\times10^{23}$\,cm$^{-2}$. 
Alternatively, the opacity change can also be parameterized by a increase in ionization with increasing flux, where $\log\xi=4.88^{+0.17}_{-0.13}$ for the 
high flux spectrum and $\log\xi=4.58^{+0.13}_{-0.12}$ for the low flux spectrum. In this scenario the column was assumed to remain constant 
with $N_{\rm H} = 3.9^{+1.4}_{-1.2}\times10^{23}$\,cm$^{-2}$. 

\begin{figure}
\begin{center}
\rotatebox{-90}{\includegraphics[height=8.5cm]{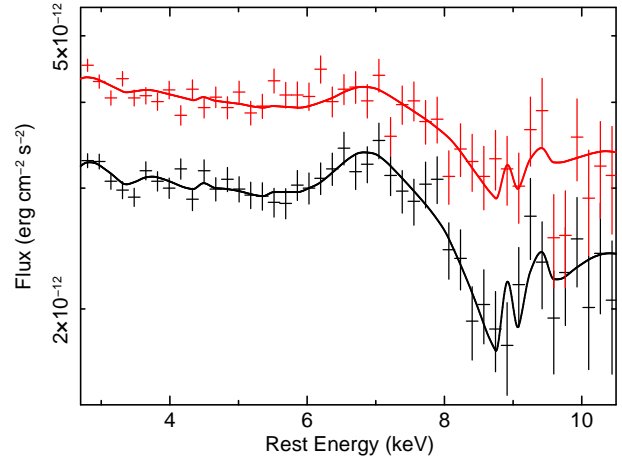}}
\end{center}
\caption{High (red) and low (black) flux spectra extracted from the 2015 observations. The best-fit \textsc{xstar} model is overlaid, which shows that the absorption trough is deeper in the low flux spectrum and can be accounted for by a factor of $\times2$ change in column density. Alternatively the change in opacity can also be modelled by variations in ionization (for a constant column), 
with the ionization increasing with X-ray flux. Overall the observations are consistent with the equivalent width of the features decreasing with increasing flux, 
as is also seen in the NLS1, IRAS\,13224-3809 \citep{Parker17}.}
\label{fig:flux}
\end{figure}

Thus on short timescales, the opacity of the wind appears to be anti-correlated with the X-ray flux. Such an effect was observed in the highly variable NLS1, IRAS\,13224-3809 
\citep{Parker17,Pinto18}, where the equivalent width of the fast iron K absorption line as well as the soft X-ray absorption features appeared to diminish with increasing flux. 
In \zw\ these changes are less drastic, as the dynamic range in 2--10\,keV flux is much smaller than in IRAS\,13224-3809. The opacity change here could either be due to 
a response in ionization of the wind to the continuum or via modest column density variations along the wind. 
In the longer term, the behaviour of the 2005 spectrum compared to 2015 appears very different. In the former despite the relatively low X-ray flux, the wind absorption is much weaker. 
However, as was recently suggested by \citet{Gallo19} for Mrk 335, the triggering of wind events could be related to coronal (ejection?) activity, which may lead to the onset of wind features during strong X-ray flares. A similar behaviour was observed in PDS 456, during a long 2013 Suzaku observation (covering a 1.5\,Ms baseline). There, the wind features emerged only following a major X-ray flare \citep{Gofford14,Matzeu17} and pre-flare, despite the relatively low X-ray flux, no Fe K absorption was present. Further, more intensive monitoring on \zw\ would be required in order to fully understand the wind variability and how it may be related to the continuum variability.


\section{Disk Wind Modeling}

In order to self consistently model the wind signatures in the \zw\ spectra, 
we utilized the radiative transfer disk wind code developed by \citet{Sim08,Sim10b}. This model creates tables of synthetic wind spectra computed for
parameterized models of smooth, steady-state 3D bi-conical winds, 
adopting the Monte Carlo ray tracing methods described by  \citet{Lucy02,Lucy03}.
The computed spectra contain both the radiation transmitted through the wind and 
reflected or scattered emission from the wind, including the iron K$\alpha$ emission.
The disk wind model thus provides a self consistent treatment of both the emission and absorption 
arising from the wind, with a physically realistic geometry, as well as computing the (non-uniform) ionization structure and velocity field 
through the flow. The model incorporates extensive 
atomic data, covering a wide range in ionization; e.g. ions from Fe\,\textsc{x-xxvi} are included as well as those from lighter elements

\begin{figure}
\begin{center}
\rotatebox{0}{\includegraphics[width=8.5cm]{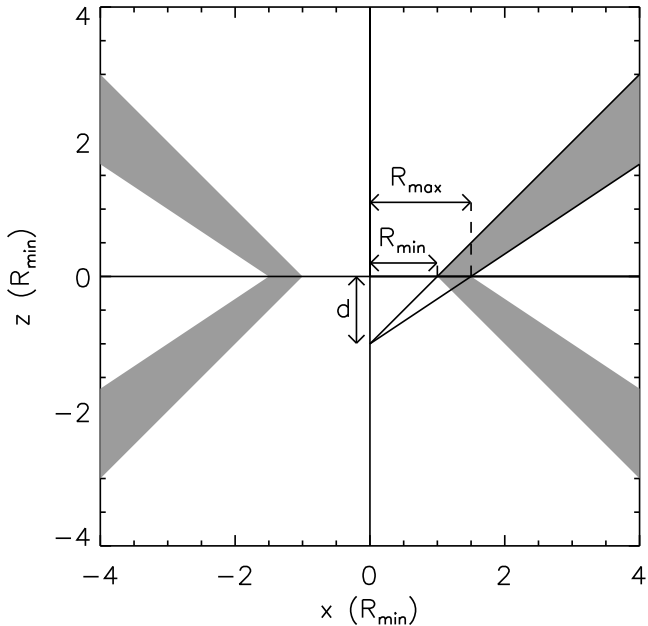}}
\end{center}
\caption{Schematic of the inner disk wind model geometry. 
The $x-$axis represents the plane of the disk and the $z-$axis 
in the polar direction, in units of $R_{\rm min}$, the minimum launch radius of the flow. 
The black hole is at the origin and  the inclination angle is measured with 
respect to the z-axis. The shaded area represents the physical extent of the outflow.
The models presented here have a minimum launch radius $R_{\rm min}=32R_{\rm g}$ along the disk plane. 
In this example, the maximum launch radius off the disk is $R_{\rm max} = 1.5R_{\rm min}$; 
this parameter sets the geometric thickness of the flow. $d$ is the distance of the focus point of the wind below the origin in units 
of $R_{\rm min}$ (here $d=1$), as indicated by the solid lines. 
Increasing $d$ makes the wind more polar and more collimated.} 
\label{fig:schematic}
\end{figure}

The \citet{Sim08} wind model has been previously employed to fit the X-ray absorption absorption profiles in several AGN, e.g. Mrk 766 (\citealt{Sim08}), 
PG\,1211+143 (\citealt{Sim10a}) and PDS\,456 (\citealt{Reeves14}), as well as the iron K$\alpha$ emission line profiles of several AGN (\citealt{Tatum12}). 
Another disk wind model, similar in geometry to the \citet{Sim08}  model, but simplified by using only He and H-like ions, was later employed by \citet{Hagino15}, 
who reproduced the Fe K absorption profiles in AGN such as PDS\,456, as well as in 1H\,0707-495 and APM\,08279+5255 (\citealt{Hagino16,Hagino17}).
Magneto hydrodynamical models have also been able to successfully model the spectra from fast outflows such as in the  
AGN PG\,1211+143 (\citealt{Fukumura15}), as well as slower winds in Galactic black hole sources, such as in GRO\,J1655-40 (\citealt{Fukumura17}). 

\subsection{The Disk Wind Parameters}

The inner wind geometry is illustrated in Figure~8 and is determined by the parameters below, which are not allowed to vary in any of the models. 
A more detailed description of the model set-up can be found in \citet{Sim08}. 

\begin{itemize}
\item Launch radius. $R_{\rm min}$ and $R_{\rm max}$ are the inner and outermost launching radii of the wind off the disk surface, which also determines the overall thickness of the wind streamline. In the model for \zw, we adopted an inner launch radius of $32 R_{\rm g}$ (where $R_{\rm g}$ is the gravitational radius), 
while we set $R_{\rm max}=1.5R_{\rm min}$ for the wind thickness. The inner wind radius was chosen as this corresponds to the 
escape radius for a wind of $v_{\infty}=0.25c$.
\item Geometry. The wind collimation and opening angle are set by the geometrical parameter $d$ (see Figure~8), 
which also determines how equatorial (or polar) the wind is.
Here $d$ is defined as the distance of the focus point of the wind below the origin in units 
of $R_{\rm min}$. In the models below, a value of $d=1$ was adopted, 
which at large radii ($R>>R_{\rm min}$) corresponds to the wind having an opening angle of $\theta=\pm45\degg$ with respect 
to the polar ($z$) axis.
\end{itemize}

\noindent In addition, the outer boundary of the disk wind simulations were set to an outer radius of $\log (R_{\rm out}/R_{\rm g})=4.5$, or $\sim3.2\times10^{4} R_{\rm g}$ ($1000R_{\rm min}$). The X-ray source is assumed to originate from a region of radius of $6R_{\rm g}$ and is centered at the origin. Special relativistic effects are accounted for in the models.

Having set up the geometric conditions of the flow, several parameters then determine the properties of the output spectra, which are described below.

\begin{itemize}
\item {Terminal velocity}. 
The terminal velocities ($v_{\infty}$) realized in the wind models are determined by
the choice of the inner wind radius ($R_{\rm min}$) and the terminal velocity parameter 
$f_{\rm v}$, which relates the terminal velocity on a wind streamline to
the escape velocity at its base, via $v_{\infty} = f_{\rm v}
\sqrt{2GM_{\rm BH}/R}$. The terminal velocity is adjusted by varying the $f_{\rm v}$ parameter, for a given launch radius (here $R_{\rm min}=32R_{\rm g}$). 
For \zw, output models were generated for 7 velocity values, ranging from $f_{\rm v}=0.5-2.0$, corresponding to terminal velocities of $0.125c-0.50c$.

\item Input Continuum. This was set to be a power-law, covering the likely range for \zw\ from $\Gamma=2.0-2.4$, over $n=3$ increments.
Note that the spectra are calculated over the $0.1-500$\,keV range.

\item Inclination angle. The observer's inclination towards the wind is defined as $\mu = \cos\theta$, where $0.025<\mu<0.975$ over 20 incremental values (with $\Delta\mu=0.05$). Here, $\theta$ is the angle between the observer's line-of-sight and the polar $z$ axis of the wind, with the disk lying in the $xy$ plane, see Figure~8.
Note that for $\mu>0.7$ (i.e. $\theta<45\degg$), the observer's line of sight does not intercept the wind, however the output spectra contain a contribution from X-ray reflection, via photons scattered off the wind. For $\mu<0.7$, the line of sight intercepts the wind, imprinting blue-shifted absorption features, while the output spectra also contain a contribution from scattered photons in emission. 

\item Mass outflow rate. This is defined by the ratio $\dot{M}=\dot{M}_{\rm out}/\dot{M}_{\rm Edd}$, where the mass outflow rate is normalized to 
the Eddington value. As a result this parameter, as well as the luminosity below, are invariant upon the black hole mass.
The grid of models here was generated covering the range $\dot{M}=0.02-0.68$, in $n=12$ increments, with equal logarithmic spacing. 
Higher values of $\dot{M}$ result in spectra with stronger emission and absorption features.

\item Ionizing X-ray luminosity. The X-ray luminosity is parameterized in the 2-10\,keV band as a percentage of the Eddington luminosity, 
where $L_{\rm X} = L_{\rm 2-10 keV}/L_{\rm Edd}$. 
The luminosity parameter sets the overall ionization of the wind and lower $L_{\rm X}$ values result in the wind being less ionized and more opaque to X-rays. 
Here, the spectral models were generated over the range of $L_{\rm X}$ of 0.025\% to 0.8\% of the Eddington luminosity, over $n=7$ increments in equal logarithmic spacing. These relatively low $L_{\rm X}$ values ensure the wind is not overly ionized so as to produce a featurless output spectrum.
\end{itemize} 

The result of the above simulations is a grid of models with $7\times3\times20\times12\times7=35280$ synthetic spectra, covering the above parameter space in 
$f_{\rm v}$, $\Gamma$, $\mu$, $\dot{M}$ and $L_{\rm X}$. The spectra are tabulated into \textsc{fits} format files and are used as multiplicative grids within 
\textsc{xspec}. Note that in the fitting procedure, interpolation is used to determine best fit parameter values and their errors if these fall in-between grid points.

\begin{figure}
\begin{center}
\rotatebox{-90}{\includegraphics[height=8.7cm]{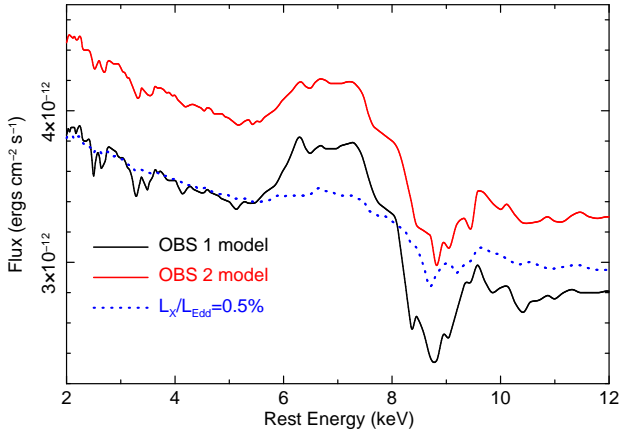}}
\end{center}
\caption{Model spectra generated with the disk wind radiative transfer code of \citet{Sim08}. The solid black and red lines correspond to the best fitting 
models to the OBS\,1 and OBS\,2 spectra, which have a mass outflow rate of 20\% of Eddington (see Table~4 for details). 
The model can reproduce Fe K profile, with a terminal velocity of $v/c=-0.28\pm0.01$. The dashed blue curve shows the effect upon the model when the 
incident $2-10$\,keV luminosity is increased from 0.14\% to 0.5\% of Eddington for OBS\,1. As a result, the profile is much weaker as the gas becomes more ionized and 
subsequently fails to account for the Fe K profile.}
\label{fig:wind}
\end{figure}

\subsection{Disk Wind Results}

The above disk wind model was then applied to the two 2015 pn spectra of \zw. 
The form of the model is simply $\textsc{tbabs}\times\textsc{diskwind}\times\textsc{powerlaw}$.
Note that the input photon index of the diskwind model is tied to the direct power-law continuum. The OBS\,1 and OBS\,2 spectra were fitted simultaneously as 
above, while all the diskwind parameters, except for the ionizing luminosity, were tied between the datasets. The power-law normalizations were  
allowed to vary independently. The results of the diskwind fits are summarized in Table~4 and produced a good fit to the data, 
with a fit statistic of $\chi_{\nu}^2=83.7/82$. An inclination angle of $\sim50\degg$ ($\mu=0.65\pm0.02$) was required, implying that the sightline directly intercepts the innermost fastest wind streamline. The terminal velocity is similar to what was found previously, with $v_{\infty}=-0.28\pm0.01c$ (corresponding to 
$f_{\rm v}=1.11\pm0.04$). 

Figure~9 shows the output spectra for the best-fit models fitted to OBS\,1 and OBS\,2, which reproduces both the 
Fe K emission and absorption well.  The best-fit mass outflow rate for these spectra (tied between the two observations) is $\dot{M}=\dot{M}_{\rm out}/\dot{M}_{\rm Edd}=0.19^{+0.16}_{-0.08}$, i.e. about 20\% of Eddington. The ionizing X-ray luminosity is $L_{\rm X}=0.14^{+0.16}_{-0.07}$\% for OBS1\, and is only marginally 
higher for OBS\,2, with $L_{\rm X}=0.22^{+0.24}_{-0.11}$\%. Overall the model prefers a solution whereby the incident $2-10$\,keV luminosity is about 0.2\% of the Eddington value. In comparison, the observed $2-10$\,keV luminosity is about $6\times10^{43}$\,erg\,s$^{-1}$, closer to 2\% of the Eddington 
luminosity, i.e. an order of magnitude higher than predicted by the model. This may suggest the wind is under-ionized compared to what one would predict
from the observed X-ray luminosity. Indeed in Figure~9 we also plot the predicted profile if the X-ray luminosity is increased in OBS\,1 to 0.5\% of Eddington (blue dotted curve); this increases the wind ionization, resulting in a much shallower profile than is observed in the data, with much weaker emission and absorption.

\begin{figure}
\begin{center}
\rotatebox{-90}{\includegraphics[height=8.7cm]{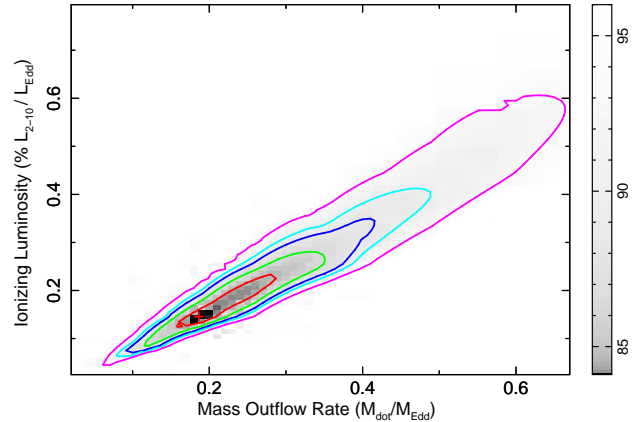}}
\end{center}
\caption{Confidence contours for the ionizing luminosity versus mass outflow rate for the wind model, where the contours represent the 68\%, 90\%, 95\%, 99\% and 
99.9\% significance levels for 2 interesting parameters. The color bar shows the $\chi^2$ space in grey scale, where darker regions have lower $\chi^2$ values; 
note the best fit is $\chi_{\nu}^2=83.7/82$. 
The mass outflow rate is constrained between 11--35\% of Eddington at 90\% confidence, while the ionizing 2--10\,keV X-ray luminosity is lower than 
the expected 2\% of Eddington value, which suggests the wind is under ionized.}
\label{fig:contour}
\end{figure}

To investigate this, we ran confidence contours between $\dot{M}$ and $L_{\rm X}$ to explore the full model parameter space and allowing the other wind parameters to vary at each grid point. Figure\,10 shows the contours for OBS\,1, at the 68\%, 90\%, 95\%, 99\% and 99.9\% confidence levels for two parameters of interest. This confirms that the mass outflow rate is constrained between 11--35\% of Eddington at the 90\% level. The direction of the contours show that as the X-ray luminosity increases, the mass outflow rate also increases to try to compensate (by increasing the column density along the flow). However, higher luminosities can still be ruled out; e.g. the 99\% upper limit is $L_{\rm X}<0.4$\%, as the flow becomes too highly ionized to reproduce the strong Fe K profile. This discrepancy between the ionizing vs observed luminosity could arise if the outflow is at least partially shielded from the X-ray continuum source, or if the 
hard X-ray continuum above 10\,keV is steeper than expected.
It may also be the case that the wind is inhomogeneous and clumpy, as opposed to the smooth wind model investigated here.
Indeed by involving a moderate micro-clumping factor of $\sim0.01$, \citet{Matthews16} can account for some of the UV line profiles in BAL QSOs, without requiring the intrinsic X-ray
luminosity to be suppressed.

\begin{deluxetable}{lcc}
\tablecaption{Disk Wind Model Results.}
\tablewidth{250pt}
\tablehead{\colhead{Parameter} & \colhead{OBS\,1} & \colhead{OBS\,2}}
\startdata
$\dot{M}_{\rm out}/\dot{M}_{\rm Edd}$$^{a}$ & $0.19^{+0.16}_{-0.08}$ & $0.19^t$\\
\% $L_{2-10}/L_{\rm Edd}$$^b$ & $0.14^{+0.16}_{-0.07}$ & $0.22^{+0.24}_{-0.11}$\\
$v_{\infty}/c$  & $-0.280\pm0.010$ & $-0.28^t$ \\
$\mu=\cos\theta$$^c$ & $0.647\pm0.015$ & $0.647^t$\\
$\Gamma$ & $2.16\pm0.03$ & $2.16^f$ \\
$L_{\rm 2-10\,keV}$$^{d}$ & 5.5 & 6.3 
\enddata
\tablenotetext{a}{Mass outflow rate in Eddington units.}
\tablenotetext{b}{Percentage ionizing ($2-10$\,keV) luminosity to Eddington luminosity.}
\tablenotetext{c}{Cosine of wind inclination, wrt the Polar axis.}
\tablenotetext{d}{Intrinsic 2--10\,keV luminosity, in units of $\times10^{43}$\,erg\,s$^{-1}$.}
\tablenotetext{t}{Denotes parameter is tied between observations.}
\label{tab:wind}
\end{deluxetable}

\section{Discussion} \label{sec:discussion}

\subsection{Kinematics of the X-ray wind}

Here we compute the energetics of the fast X-ray wind for the three different wind models, comparing both the \textsc{xstar} and P-Cygni cases with the diskwind model above. 
Following the methodology of \citet{Nardini15}, we adopt the mass outflow rate in the form:-

\begin{equation}
\dot{M}_{\rm out} = \Omega \mu m_{\rm p} N_{\rm H} R_{\rm in} v
\end{equation}

\noindent where $\Omega$ is the overall wind solid angle, $\mu m_{\rm p}$ is the mean baryonic mass per particle ($\mu\sim1.3$ for Solar abundances), $N_{\rm H}$ is the hydrogen column density, $v$ is the wind (terminal) velocity and $R_{\rm in}$ is the inner launch radius of wind.
For the \textsc{xstar} model we adopt $\Omega/4\pi=0.5$ (consistent with the ratio between the absorbed vs emitted flux), while the P-Cygni model (spherical wind) implicitly assumes $\Omega/4\pi=1$. For the column density, for the \textsc{xstar} model we calculated the mean value over the 
four observations in Table~3, of $N_{\rm H}=4.7\pm1.1\times10^{23}$\,cm$^{-2}$, while for the P-Cygni model, we took the average of the two 2015 observations in Section~3.2, of $N_{\rm H}=5.0\pm0.9\times10^{23}$\,cm$^{-2}$. 

The wind launch radius was set to the escape radius from the black hole, i.e. $R_{\rm in}=2GM_{\rm BH}/v^2$, which gives the smallest inner radius of the wind and thus a more conservative estimate of the mass outflow rate. The mass outflow rate, normalized to the Eddington rate where $\dot{M}_{\rm Edd}=4\pi GM_{\rm BH} m_{\rm p}/\sigma_{\rm T} \eta c$, is then:-

\begin{equation}
\dot{M}=\frac{\dot{M}_{\rm out}}{\dot{M}_{\rm Edd}} = 2 \frac{\Omega}{4\pi} \mu N_{\rm H} \sigma_{\rm T} \eta \left(\frac{v}{c}\right)^{-1}
\end{equation}

\noindent where we adopt $\eta=0.1$ for the accretion efficiency and $\sigma_{\rm T}$ is the Thomson cross section. 
Subsequently the wind kinetic power, $L_{\rm k}$, normalized to the Eddington luminosity, is:-

\begin{equation}
\dot{E}=\frac{L_{\rm k}}{L_{\rm Edd}} = \frac{1}{2\eta} \frac{\dot{M}_{\rm out}}{\dot{M}_{\rm Edd}} \left(\frac{v}{c}\right)^{2} = \frac{\Omega}{4\pi} \mu N_{\rm H} \sigma_{\rm T} \frac{v}{c}
\end{equation}

\noindent The wind momentum rate (thrust) of the wind is $\dot{p}_{\rm out}=\dot{M}_{\rm out} v$, while that of the radiation field is $\dot{p}_{\rm Edd}=L_{\rm Edd}/c$ (where for \zw\ $L_{\rm bol} \sim L_{\rm Edd}$). Thus:-

\begin{equation}
\dot{p}=\frac{\dot{p}_{\rm out}}{\dot{p}_{\rm Edd}} = \frac{1}{\eta} \frac{\dot{M}_{\rm out}}{\dot{M}_{\rm Edd}} \frac{v}{c} = 2 \mu \frac{\Omega}{4\pi} N_{\rm H} \sigma_{T} \sim \tau
\end{equation}

\noindent and $\tau$ is the optical depth to Compton scattering. Thus the inner wind is approximately momentum conserving against the radiation field in the single photon scattering limit, where $\tau\sim1$ (e.g. \citealt{KP03}).

We calculated the above values of $\dot{M}$, $\dot{E}$ and $\dot{p}$ for the both the \textsc{xstar} and P-Cygni models and we used the best-fit value of 
$\dot{M}$ for the diskwind model to calculate the corresponding values of $\dot{E}$ and $\dot{p}$. Table~5 shows the resulting values, while we also give the absolute values of 
the mass outflow rate ($\dot{M}_{\rm out}$) and kinetic luminosity ($L_{\rm k}$), for a black hole mass of $M_{\rm BH}=2.8^{+0.6}_{-0.7}\times10^7$\,M$_{\odot}$ and 
a corresponding Eddington luminosity of $L_{\rm Edd}=3.5\times10^{45}$\,erg\,s$^{-1}$. 

Overall, the derived mass outflow rate is between $15-25$\% of Eddington, corresponding to $\sim0.1-0.15$\,M$_{\odot}$\,yr$^{-1}$, while the wind kinetic power ranges 
between $5-15$\% of Eddington (or $L_{\rm k}=2-5\times10^{44}$\,erg\,s$^{-1}$), which is potentially significant for mechanical feedback on larger scales (\citealt{HopkinsElvis10}). These values are typical of those found in ultra fast outflows in other AGN (\citealt{Tombesi13,Gofford15}). 
The wind momentum rate is of the order unity or just below ($\dot{p}\sim0.4-0.9$), which is also in agreement with other ultra fast outflows; e.g. see Figure 4, \citet{Tombesi13}. 

Reassuringly, the outflow rate and kinetic power obtained between the three models are consistent within the uncertainties, despite of any differences in physical construction between them. Of the three models, the \textsc{xstar} model has the most complete atomic physics, but is the least self consistent geometrically, 
as the absorption is computed from a one dimensional slab, 
although the comparison between the total emission vs absorption does make it possible to estimate the overall covering fraction of the gas. 
Both the P-Cygni and disk wind models have the advantage of being 
physically motivated wind models, where the subsequent velocity profiles (in emission and absorption) are self 
consistently calculated for a given terminal velocity over all solid angles, where the former assumes a spherical wind and the latter a bi-conical outflow. 
The disk wind model has the further advantage in that the outflow parameters derived are independent of the black hole mass, as the output spectra are invariant upon this parameter for a realistic range of AGN black hole masses.
Of all three models, the values for the P-Cygni model lie the upper end of the range, due to its spherical geometry and the higher terminal velocity obtained ($v_{\infty}=0.35^{+0.03}_{-0.04}c$), which could be considered to be the least conservative scenario. 

\subsubsection{Comparison with the wind in PDS 456}

We also compare the wind properties of \zw\ with those obtained from the prototype disk wind quasar, PDS\,456. For illustration, Figure\,11 shows the Fe K wind profile of \zw\ compared to the spectrum consisting of the third and fourth observations from the 2013--2014 \xmm\ campaign on PDS\,456. This is the same spectrum which was analyzed by \citet{Nardini15}, where a broad Fe K P-Cygni profile was first found in this QSO. 
The profiles of PDS\,456 and \zw\ are remarkably similar, as both show an excess in emission above the continuum between 6--8\,keV, while a deep broad absorption trough is present above 8\,keV in both quasars. It is noticeable that the blue-shift of the absorption trough is slightly larger in PDS\,456, where $v/c=0.29\pm0.01$ from the absorption line centroid, although we note that the outflow velocity in PDS\,456 has been observed to vary from $0.25-0.35c$ over all the X-ray observations (2001--2017) to date (\citealt{Matzeu17}). 

\begin{figure}
\begin{center}
\rotatebox{-90}{\includegraphics[height=8.7cm]{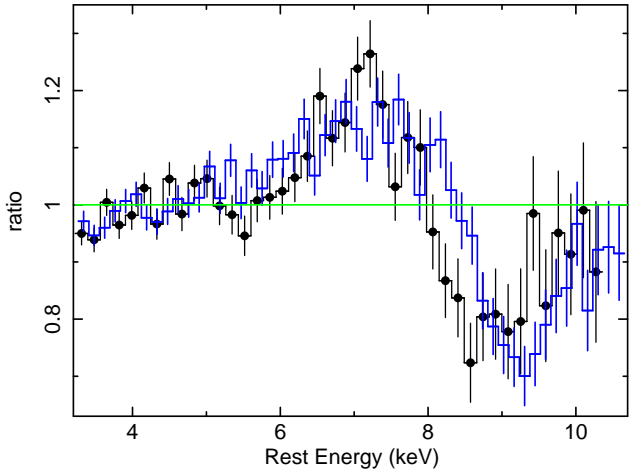}}
\end{center}
\caption{Comparison between the Fe K wind profiles in \zw\ and the quasar, PDS\,456. The black circles show the \zw\ 2015 OBS\,1 profile, while the blue points 
show the pn profile of PDS\,456, from the third and fourth observations of the 2013--2014 \xmm\ campaign (Nardini et al. 2015). Both AGN are plotted in their respective rest-frames ($z=0.184$ for PDS\,456, $z=0.0611$ for \zw) and are shown as a ratio against a simple absorbed power-law continuum. The profiles show strong similarities; both require broadened emission centered near to 7 keV as well as a broad, blue-shifted absorption trough above 8\,keV, of similar optical depth. The blue-shift of the 
2013 PDS\,456 profile is slightly higher than in \zw, suggesting a somewhat higher outflow velocity. The wind energetics are similar for both AGN, with the mass outflow rate and kinetic power typically of 10\% Eddington (see Section 6.1). Overall, \zw\ may represent a lower mass analogue of the powerful disk wind in PDS\,456.}
\label{fig:pds456}
\end{figure}

From analyzing all five \xmm\ and {\it NuSTAR} observations of PDS\,456 in 2013--2014, Nardini et al. (2015) obtained an average column density of 
$N_{\rm H}=6.9^{+0.6}_{-1.2}\times10^{23}$\,cm$^{-2}$ for the Fe K wind, while the profile was consistent with a minimum covering of $\Omega=2\pi$\,sr. 
Adopting these values and setting the inner wind radius equal to the escape radius for consistency, then for PDS\,456 $\dot{M}\sim0.2$ (or $\sim5$\,M$_{\odot}$\,yr$^{-1}$), while $\dot{E}\sim0.1$ (or $L_{\rm k}\sim10^{46}$\,erg\,s$^{-1}$) and $\dot{p}\sim1$. In terms of their Eddington values, the outflow energetics are very similar between both \zw\ and 
PDS\,456, with the higher absolute values in PDS\,456 being due to its larger black hole mass of $10^{9}$\,M$_{\odot}$. 
Thus the wind in \zw\ may be a lower mass analogue of the one in PDS\,456, where in both AGN, the high accretion rate close to Eddington likely creates favourable conditions for driving a fast disk wind.

\begin{deluxetable}{lcccc}
\tablecaption{Derived Outflow Energetics for \zw.}
\tablewidth{250pt}
\tablehead{& \colhead{xstar} & \colhead{diskwind} & \colhead{P-Cygni} & \colhead{CO}}
\startdata
$\dot{M}^{a}$ & $0.15\pm0.04$ & $0.19^{+0.16}_{-0.08}$ & $0.25\pm0.05$ & --\\
& ($0.09\pm0.04$) & ($0.11^{+0.15}_{-0.06}$) & ($0.15\pm0.06$) & ($<140$)\\
$\dot{E}^b$ & $0.054\pm0.013$ & $0.08^{+0.07}_{-0.03}$ & $0.15\pm0.04$ & $<3\times10^{-3}$\\
& ($2\times10^{44}$) & ($3\times10^{44}$) & ($5\times10^{44}$) & ($<1\times10^{43}$)\\
$\dot{p}^{c}$  & $0.4\pm0.1$ & $0.5^{+0.4}_{-0.2}$ & $0.9\pm0.2$ & $<4$ 
\enddata
\tablenotetext{a}{Mass outflow rate in Eddington units. Note absolute values are given in parenthesis in units of Solar masses per year.}
\tablenotetext{b}{Outflow kinetic power in Eddington units. Absolute values are given in parenthesis in units of ergs\,s$^{-1}$.}
\tablenotetext{c}{Outflow momentum rate in Eddington units.}
\label{tab:energetics}
\end{deluxetable}

\subsection{Comparison with the Molecular gas Outflows}

The energetics for the X-ray wind are now compared to observations of the molecular gas in \zw, through the CO($1-0$) line, to 
ascertain the mechanical effect of the wind on the large scale star-forming ISM gas.
Observations of \zw\ were made with the IRAM PdBI in 2010, which spatially and kinematically resolved the CO emission on $\sim$kpc scales and the results 
were reported in \citet{Cicone14}. Unlike the clear signature of outflow for the X-ray wind, the CO observations show no clear evidence of any large scale molecular outflow. 
The continuum subtracted CO emission line profile (see \citealt{Cicone14}, Figure 7) is narrow with no apparent blue-shift or broad-wings, yielding an upper limit to its velocity width of  $<500$\,km\,s$^{-1}$ (at FWZI). This is also consistent with the lack of any blue-shift in either emission or absorption 
in \zw\ from the OH profiles as measured by {\it Herschel}-PACS (\citealt{Veilleux13}). The CO velocity map does show clear evidence of rotation, via double-peaked emission and which is likely associated with a disk of molecular gas within the host galaxy. 

\citet{Cicone14} subsequently derived an upper limit of $140\,M_{\odot}$\,yr$^{-1}$ for the total molecular mass outflow rate in \zw, after adopting a conservative upper limits of $v<500$\,km\,s$^{-1}$ from the CO profile and $r<500$\,pc from the spatial extent of the narrow core. For comparison with the X-ray wind, we then computed the upper limits to both the kinetic power and momentum rate for any molecular outflow, which are also reported in Table~5. 
The upper limit to the kinetic power for $v<500$\,km\,s$^{-1}$ is subsequently 
$L_{\rm K}<1\times10^{43}$\,erg\,s$^{-1}$, corresponding to $L_{\rm K}/L_{\rm Edd}<3\times10^{-3}$. This is more than an order of magnitude lower than the kinetic power of the 
X-ray wind, which gave $L_{\rm K}=2-5\times10^{44}$\,erg\,s$^{-1}$.
On the other hand the momentum rate for the molecular gas is $\dot{p}/\dot{p}_{\rm Edd}<4.0$, which is consistent with the 
outflow being momentum rather than energy conserving on large scales.


\begin{figure}
\begin{center}
\rotatebox{-90}{\includegraphics[height=8.7cm]{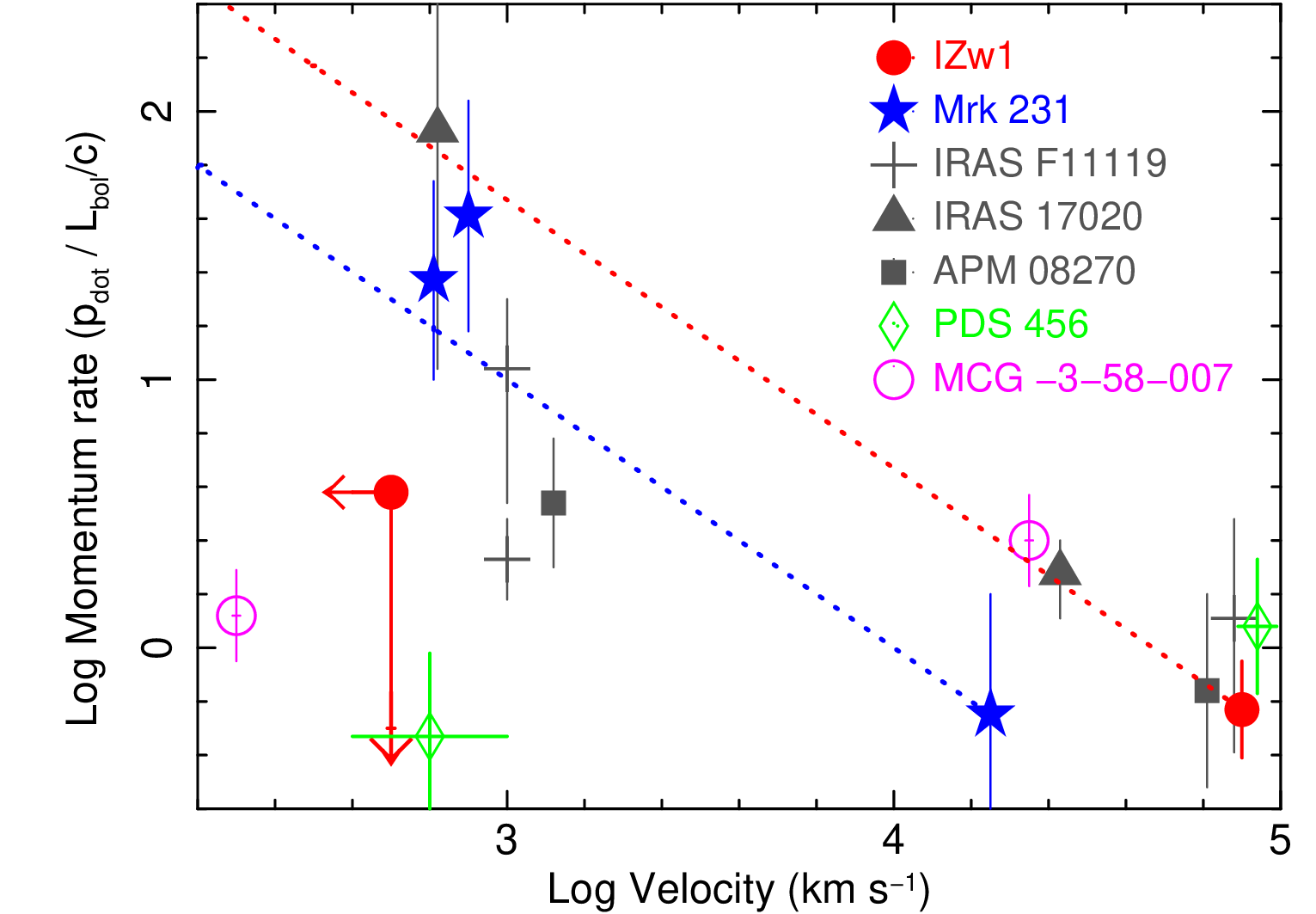}}
\end{center}
\caption{Comparison between momentum rates (normalized by $L_{\rm bol}/c$) and outflow velocity for the six ultra fast outflows with reported measurements of molecular outflows. Data points as marked correspond to \zw\ (\citealt{Cicone14}; this paper), Mrk\,231 (\citealt{Feruglio15}), IRAS\,F$11119+3257$ (\citealt{Tombesi15,Veilleux17}), IRAS\,$17020+4544$ (\citealt{Longinotti15,Longinotti18}), APM~$08279+5255$ (\citealt{Feruglio17}), PDS\,456 (\citealt{Bischetti19}) and MCG--03-58-007  \citep{Braito18,Sirressi19}.  
Note that for Mrk\,231, both CO components are plotted separately, as well as the separate CO and OH measurements of IRAS\,F$11119+3257$.
The dotted lines correspond to the predicted momentum load factors in \zw\ (red line) and Mrk\,231 (blue line) for the energy conserving scenario. The molecular wind in Mrk\,231 is consistent with being energy driven, in contrast to \zw, PDS\,456 and MCG--03-58-007, where the momentum rate of the molecular gas lies about two orders of magnitude below the energy conserving relation.}
\label{fig:pdot}
\end{figure}

The outflow in \zw\ can also be compared with other AGN where both an ultra fast disk wind and a kpc scale molecular outflow co-exist.  In Figure~12, we plot the momentum rates of the X-ray versus molecular phases of the outflows against wind velocity for the five AGN which have been reported to host both an ultra fast disk wind and a large scale molecular outflow.  The first reported examples were in the ULIRGs / type II QSOs, Mrk\,231 (\citealt{Feruglio15}) and IRAS\,F$11119+3257$ (\citealt{Tombesi15}), 
where the detection of an energy conserving molecular outflow implied that the kinetic power of the fast wind was effectively transferred out to larger scale gas.
However, even from the small sample plotted in Figure~12, it is apparent that not all of the large scale outflows lie on the predicted relation for an energy driven wind. 
Indeed, only Mrk\,231 and the NLS1, IRAS\,$17020+4544$ (\citealt{Longinotti15,Longinotti18}), are consistent with the scenario whereby the black hole wind drives an energy conserving 
molecular outflow out to large scales. In complete contrast, an energy conserving wind can be clearly ruled out in \zw, as the momentum rate for the molecular gas lies nearly two orders of magnitude below the predicted value for an energy-driven outflow.  A similar scenario also applies to  the high redshift QSO, APM~$08279+5255$ (\citealt{Feruglio17}), where the measured  momentum boost is lower than the prediction for  an energy-conserving wind.  Even the original case of IRAS\,F$11119+3257$ now appears more complex, as recent ALMA CO measurements (\citealt{Veilleux17}) suggest that the momentum boost is lower than that originally derived from the Herschel OH profile in  \citet{Tombesi15} and which was also further noted by \citet{NardiniZubovas18}. 

These results suggest a range of efficiencies in transferring the kinetic energy of the inner wind out to the large-scale molecular component. 
This is consistent with the recent analysis of \citet{Mizumoto19}, who analyzed a small X-ray sample of 8 AGN (including \zw), selected from the \citet{Cicone14} CO sample. \citet{Mizumoto19} inferred that the energy transfer rate, from the black hole wind to the molecular gas, spanned a wide range of efficiency from 
$\epsilon\sim7\times 10^{-3} - 1$ (where $\epsilon=1$ corresponds to a perfectly energy conserving outflow). 
Similarly, the AGN outflows compiled by \citet{Fiore17} show a wide range of momentum load factors, with about half of the AGN 
receiving a boost of at least $\times10$ in the molecular outflow component compared to the X-ray wind.

We also caution that comparing the energetics of the nuclear and larger scale outflows is challenging, especially when the  host is a powerful ultra luminous infrared galaxy (ULIRG), which are generally characterized by a high star formation rate. In these objects, powerful galactic-scale outflows are common and could involve different gas phases (Cicone 2018).  Nonetheless, the energetics of the molecular outflows seen in Mrk~231 and IRAS~$17020+4544$  (where $L_{\rm k}>10^{44}$\,erg\,s$^{-1}$) are too large to be solely explained by the star formation activity. The large momentum boost  ($\dot{p}_{\rm CO}/\dot{p}_{\rm X}>10$) measured for both these outflows suggest that the mechanical energy of the inner AGN disk wind is efficiently carried to the large scale outflows.  

In \zw, like in PDS\,456, the AGN emission dominates with respect to the star forming activity.  Interesting, from very recent {\it ALMA} observations, \citet{Bischetti19} for the first time resolved the large scale molecular outflow in CO from PDS\,456. These observations revealed a complex and clumpy outflow, extending up to 5\,kpc from the nucleus. 
The total molecular mass outflow rate measured in PDS\,456 is $180-760$\,M$_{\odot}$\,yr$^{-1}$, with a corresponding momentum rate of 
$7.8-32\times10^{35}$\,dyne. Compared to the bolometric luminosity of PDS\,456 of $10^{47}$\,erg\,s$^{-1}$, then $\dot{p}_{\rm CO}/\dot{p}_{\rm rad}\ls 1$; this is clearly consistent with a momentum rather than energy conserving outflow.
Indeed the very sensitive {\it ALMA} observations are very stringent in PDS\,456 and place the quasar in the same region of the momentum rate versus velocity diagram as \zw (see Figure 12). \citet{Bischetti19} suggest that part of the large scale outflow in PDS\,456 could plausibly be in the form of ionized instead of molecular gas, thereby 
leading to the total mass outflow rate to be under-estimated. Alternatively, the conditions for a large scale energy conserving outflow (\citealt{ZubovasKing12}) could break down in such a luminous quasar. 

Another powerful disk wind was recently discovered in the nearby ($z=0.031462$), luminous Seyfert 2 galaxy, MCG--03-58-007 (\citealt{Braito18,Matzeu19}). This AGN appears similar to both \zw\ and PDS\,456 in its X-ray characteristics, while the AGN has a similar bolometric luminosity to \zw, of $L_{\rm bol}\sim3\times10^{45}$\,erg\,s$^{-1}$. 
In this object, the disk wind has a kinetic power of $\sim 10$\% of $L_{\mathrm {bol}}$, while its host galaxy has only a modest star formation rate ($\sim 10M_\odot $\,yr$^{-1}$; \citealt{Oi10}).  
The subsequent analysis of a recent ALMA observation of this AGN \citep{Sirressi19}, shows that it hosts a rather weak kpc scale outflow in CO, consistent with a momentum driven wind with $\dot{p}_{\rm CO}/\dot{p}_{\rm X}\sim1$  and where the energy efficiency factor between the molecular and X-ray wind is remarkably low 
($\epsilon<10^{-3}$), similar to both PDS\,456 and \zw. 
Further CO measurements of AGN with fast X-ray winds (and vice versa) 
are now clearly required to establish how efficient or not the black hole winds are at transferring their mechanical energy out to gas at kpc scales and thus their overall 
role in large scale feedback.

\section{Acknowledgements}

We would like to thank Stuart Sim for the use of his disk wind radiative transfer code used in this paper as well as Michele Costa for assistance in running the disk wind models.
JR acknowledges financial support through grants 
NNX17AC38G, NNX17AD56G and HST-GO-14477.001-A. 
Based on observations obtained with XMM-Newton, an ESA science mission with instruments and contributions directly funded by ESA Member States and NASA.


\begin{thebibliography}

\bibitem[Behar et al.(2010)]{Behar10} Behar, E., Kaspi, S., Reeves, J., et al.\ 2010, \apj, 712, 26 
\bibitem[Bianchi et al.(2007)]{Bianchi07} Bianchi, S., Guainazzi, M., Matt, G., \& Fonseca Bonilla, N.\ 2007, \aap, 467, L19
\bibitem[Bischetti et al.(2019)]{Bischetti19} Bischetti, M., Piconcelli, E., Feruglio, C., et al.\ 2019, A\&A, 628, A118
\bibitem[Boller et al.(1996)]{Boller96} Boller, T., Brandt, W.~N., \& Fink, H.\ 1996, \aap, 305, 53 
\bibitem[Braito et al.(2018)]{Braito18} Braito, V., Reeves, J.~N., Matzeu, G.~A., et al.\ 2018, \mnras, 479, 3592 
\bibitem[Brightman \& Nandra(2011)]{BrightmanNandra11} Brightman, M., \& Nandra, K.\ 2011, \mnras, 413, 1206 
\bibitem[Chartas et al.(2002)]{Chartas02}Chartas, G., Brandt, W.~N., Gallagher, S.~C., \& Garmire, G.~P.\ 2002, \apj, 579, 169

\bibitem[Chartas et al.(2009)]{Chartas09} Chartas, G., Saez, C., Brandt, W.~N., Giustini, M., \& Garmire, G.~P.\ 2009, \apj, 706, 644
\bibitem[Cicone et al.(2014)]{Cicone14} Cicone, C., Maiolino, R., Sturm, E., et al.\ 2014, \aap, 562, A21 
\bibitem[Cicone et al.(2015)]{Cicone15} Cicone, C., Maiolino, R., Gallerani, S., et al.\ 2015, \aap, 574, A14 
\bibitem[Costantini et al.(2007)]{Costantini07} Costantini, E., Gallo, L.~C., Brandt, W.~N., Fabian, A.~C., \& Boller, T.\ 2007, \mnras, 378, 873 
\bibitem[Di Matteo et al.(2005)]{DiMatteo05} Di Matteo, T., Springel, V., \& Hernquist, L.\ 2005, \nat, 433, 604
\bibitem[Done et al.(2007)]{Done07} Done, C., Sobolewska, M.~A., Gierli{\'n}ski, M., \& Schurch, N.~J.\ 2007, \mnras, 374, L15 
\bibitem[Fabian(1999)]{Fabian99} Fabian, A. C., 1999, MNRAS, 308, L39

\bibitem[Faucher-Gigu{\`e}re \& Quataert(2012)]{FaucherQuataert} Faucher-Gigu{\`e}re, C.-A., \& Quataert, E.\ 2012, \mnras, 425, 605 
\bibitem[Ferrarese \& Merritt(2000)]{FerrareseMerritt00}Ferrarese L., Merritt D., 2000, ApJ, 539, 9
\bibitem[Feruglio et al.(2010)]{Feruglio10} Feruglio, C., Maiolino, R., Piconcelli, E., et al.\ 2010, \aap, 518, L155 
\bibitem[Feruglio et al.(2015)]{Feruglio15}Feruglio C., et al., 2015, A\&A, 583, 99
\bibitem[Feruglio et al.(2017)]{Feruglio17} Feruglio, C., Ferrara, A., Bischetti, M., et al.\ 2017, \aap, 608, A30 
\bibitem[Fiore et al.(2017)]{Fiore17} Fiore, F., Feruglio, C., Shankar, F., et al.\ 2017, \aap, 601, A143 
\bibitem[Fukumura et al.(2010)]{Fukumura10} Fukumura, K., Kazanas, D., Contopoulos, I., \& Behar, E.\ 2010, \apjl, 723, L228
\bibitem[Fukumura et al.(2015)]{Fukumura15} Fukumura, K., Tombesi, F., Kazanas, D., et al.\ 2015, \apj, 805, 17 
\bibitem[Fukumura et al.(2017)]{Fukumura17} Fukumura, K., Kazanas, D., Shrader, C., et al.\ 2017, Nature Astronomy, 1, 0062 
\bibitem[Gallo et al.(2004)]{Gallo04}Gallo, L.~C., Boller, T., Brandt, W.~N., Fabian, A.~C., \& Vaughan, S.\ 2004, \aap, 417, 29 
\bibitem[Gallo et al.(2007)]{Gallo07}   Gallo, L.~C., Brandt, W.~N., Costantini, E., et al.\ 2007, \mnras, 377, 391
\bibitem[Gallo et al.(2019)]{Gallo19} Gallo, L.~C., Gonzalez, A.~G., Waddell, S.~G.~H., et al.\ 2019, \mnras, 484, 4287
\bibitem[Garc{\'{\i}}a et al.(2013)]{Garcia13} Garc{\'{\i}}a, J., Dauser, T., Reynolds, C.~S., et al.\ 2013, \apj, 768, 146
\bibitem[Gebhardt(2000)]{Gebhardt00}Gebhardt K., 2000, ApJ, 539, 13

\bibitem[Gofford et al.(2013)]{Gofford13}Gofford J., Reeves J. N., Tombesi F., et al., 2013, MNRAS, 430, 60

\bibitem[Gofford et al.(2014)]{Gofford14} Gofford, J., Reeves, J.~N., Braito, V., et al.\ 2014, \apj, 784, 77 
\bibitem[Gofford et al.(2015)]{Gofford15} Gofford, J., Reeves, J.~N., McLaughlin, D.~E., et al.\ 2015, \mnras, 451, 4169 
\bibitem[Grevesse \& Sauval(1998)]{GrevesseSauval98}Grevesse N., Sauval A. J., 1998, SSRv, 85, 161

\bibitem[Hagino et al.(2015)]{Hagino15} Hagino, K., Odaka, H., Done, C., et al.\ 2015, \mnras, 446, 663 
\bibitem[Hagino et al.(2016)]{Hagino16} Hagino, K., Odaka, H., Done, C., et al.\ 2016, \mnras, 461, 3954 
\bibitem[Hagino et al.(2017)]{Hagino17} Hagino, K., Done, C., Odaka, H., Watanabe, S., \& Takahashi, T.\ 2017, \mnras, 468, 1442 
 
\bibitem[Hamann et al.(2018)]{Hamann18} Hamann, F., Chartas, G., Reeves, J., \& Nardini, E.\ 2018, \mnras, 476, 943

\bibitem[Hopkins \& Elvis(2010)]{HopkinsElvis10}Hopkins P. F., Elvis M., 2010, MNRAS, 401, 7
\bibitem[Ikeda et al.(2009)]{Ikeda09} Ikeda, S., Awaki, H., \& Terashima, Y.\ 2009, \apj, 692, 608 
\bibitem[Iwasawa \& Taniguchi(1993)]{IwasawaTaniguchi93} Iwasawa, K., \& Taniguchi, Y.\ 1993, \apjl, 413, L15 
\bibitem[Kalberla et al.(2005)]{Kalberla05} Kalberla, P.~M.~W., Burton, W.~B., Hartmann, D., et al.\ 2005, \aap, 440, 775
\bibitem[Kallman et al.(2004)]{Kallman04} Kallman, T. R., Palmeri, P., Bautista, M. A., Mendoza, C., 
\& Krolik, J. H., 2004, ApJS, 155, 675

\bibitem[King(2003)]{King03} King, A. R., 2003, ApJ, 596, L27

\bibitem[King \& Pounds(2003)]{KP03} King, A. R., \& Pounds, K. A., 2003, MNRAS, 345, 657

\bibitem[King(2010)]{King10}King A. R., 2010, MNRAS, 402, 1516
\bibitem[King \& Pounds(2015)]{KingPounds15} King, A., \& Pounds, K.\ 2015, \araa, 53, 115 
\bibitem[Kosec et al.(2018)]{Kosec18} Kosec, P., Buisson, D.~J.~K., Parker, M.~L., et al.\ 2018, \mnras, 481, 947 \bibitem[Leighly(1999)]{Leighly99} Leighly, K.~M.\ 1999, The Astrophysical Journal Supplement Series, 125, 317
\bibitem[Longinotti et al.(2015)]{Longinotti15} Longinotti, A.~L., Krongold, Y., Guainazzi, M., et al.\ 2015, \apjl, 813, L39
 \bibitem[Longinotti et al.(2018)]{Longinotti18} Longinotti, A.~L., Vega, O., Krongold, Y., et al.\ 2018, \apjl, 867, L11 
\bibitem[Lucy(2002)]{Lucy02} Lucy, L.~B.\ 2002, \aap, 384, 725 
\bibitem[Lucy(2003)]{Lucy03} Lucy, L.~B.\ 2003, \aap, 403, 261 
\bibitem[Maiolino et al.(2012)]{Maiolino12} Maiolino, R., Gallerani, S., Neri, R., et al.\ 2012, \mnras, 425, L66 
\bibitem[Matthews et al.(2016)]{Matthews16} Matthews, J.~H., Knigge, C., Long, K.~S., et al.\ 2016, \mnras, 458, 293
\bibitem[Matzeu et al.(2016)]{Matzeu16} Matzeu, G.~A., Reeves, J.~N., Nardini, E., et al.\ 2016, \mnras, 458, 1311 
\bibitem[Matzeu et al.(2017)]{Matzeu17} Matzeu, G.~A., Reeves, J.~N., Braito, V., et al.\ 2017, \mnras, 472, L15 
\bibitem[Matzeu et al.(2019)]{Matzeu19} Matzeu, G.~A., Braito, V., Reeves, J.~N., et al.\ 2019, \mnras, 483, 2836 
\bibitem[Mizumoto et al.(2019)]{Mizumoto19} Mizumoto, M., Izumi, T., \& Kohno, K.\ 2019, \apj, 871, 156 
\bibitem[Murphy \& Yaqoob(2009)]{MurphyYaqoob09} Murphy, K.~D., \& Yaqoob, T.\ 2009, \mnras, 397, 1549 
\bibitem[Nandra et al.(1997)]{Nandra97} Nandra, K., George, I.~M., Mushotzky, R.~F., Turner, T.~J., \& Yaqoob, T.\ 1997, \apjl, 488, L91 
\bibitem[Nandra et al.(2007)]{Nandra07} Nandra, K., O'Neill, P.~M., George, I.~M., \& Reeves, J.~N.\ 2007, \mnras, 382, 194 
\bibitem[Nardini et al.(2015)]{Nardini15} Nardini, E., Reeves, J.~N., Gofford, J., et al.\ 2015, Science, 347, 860
\bibitem[Nardini \& Zubovas(2018)]{NardiniZubovas18} Nardini, E., \& Zubovas, K.\ 2018, \mnras, 478, 2274 
\bibitem[Oi et al.(2010)]{Oi10} Oi, N., Imanishi, M., \& Imase, K.\ 2010, \pasj, 62, 1509 
\bibitem[Osterbrock \& Pogge(1985)]{OsterbrockPogge85} Osterbrock, D.~E., \& Pogge, R.~W.\ 1985, \apj, 297, 166 
\bibitem[Page et al.(2005)]{Page05} Page, K.~L., Reeves, J.~N., O'Brien, P.~T., \& Turner, M.~J.~L.\ 2005, \mnras, 364, 195 
\bibitem[Parker et al.(2017)]{Parker17} Parker, M.~L., Alston, W.~N., Buisson, D.~J.~K., et al.\ 2017, \mnras, 469, 1553 
\bibitem[Parker et al.(2018)]{Parker18} Parker, M.~L., Reeves, J.~N., Matzeu, G.~A., Buisson, D.~J.~K., \& Fabian, A.~C.\ 2018, \mnras, 474, 108
\bibitem[Pinto et al.(2018)]{Pinto18} Pinto, C., Alston, W., Parker, M.~L., et al.\ 2018, \mnras, 476, 1021 
\bibitem[Porquet et al.(2004)]{Porquet04} Porquet, D., Reeves, J.~N., O'Brien, P., et al.\ 2004, \aap, 422, 85.
\bibitem[Pounds et al.(2003)]{Pounds03} Pounds, K.~A., Reeves, J.~N., King, A.~R., et al.\ 2003, \mnras, 345, 705 

\bibitem[Proga \& Kallman(2004)]{ProgaKallman04}Proga D., Kallman T. R., 2004, ApJ, 616, 688
\bibitem[Sargent(1968)]{Sargent68} Sargent, W.~L.~W.\ 1968, \apjl, 152, L31 
\bibitem[Reeves \& Turner(2000)]{ReevesTurner00} Reeves, J.~N., \& Turner, M.~J.~L.\ 2000, \mnras, 316, 234.
\bibitem[Reeves et al.(2000)]{Reeves00} Reeves, J.~N., O'Brien, P.~T., Vaughan, S., et al.\ 2000, \mnras, 312, L17

\bibitem[Reeves et al.(2003)]{Reeves03} Reeves, J.~N., O'Brien, P.~T., \& Ward, M.~J.\ 2003, \apjl, 593, L65

\bibitem[Reeves et al.(2009)]{Reeves09} Reeves, J.~N., O'Brien, P.~T., Braito, V., et al.\ 2009, \apj, 701, 493

\bibitem[Reeves et al.(2014)]{Reeves14} Reeves, J.~N., Braito, V., Gofford, J., et al.\ 2014, \apj, 780, 45 


\bibitem[Saez et al.(2009)]{Saez09} Saez, C., Chartas, G., \& Brandt, W.~N.\ 2009, \apj, 697, 194

\bibitem[Saez \& Chartas(2011)]{SaezChartas11} Saez, C., \& Chartas, G.\ 2011, \apj, 737, 91
\bibitem[Schmidt \& Green(1983)]{SchmidtGreen83} Schmidt, M., \& Green, R.~F.\ 1983, \apj, 269, 352 
\bibitem[Scott et al.(2004)]{Scott04} Scott, J.~E., Kriss, G.~A., Brotherton, M., et al.\ 2004, \apj, 615, 135 
\bibitem[Silk \& Rees(1998)]{SilkRees98} Silk, J., \& Rees, M. J., 1998, A\&A, 331, L1
\bibitem[Silva et al.(2018)]{Silva18} Silva, C.~V., Costantini, E., Giustini, M., et al.\ 2018, \mnras, 480, 2334 
\bibitem[Sim et al.(2008)]{Sim08}Sim, S. A., Long, K.S., Miller, L., \& Turner, T.J., 2008, MNRAS, 388, 611
\bibitem[Sim et al.(2010a)]{Sim10a} Sim, S.~A., Miller, L., Long, K.~S., Turner, T.~J., \& Reeves, J.~N.\ 2010, \mnras, 404, 1369 
\bibitem[Sim et al.(2010b)]{Sim10b}Sim S. A., Proga D., Miller L., Long K. S., Turner T. J., 2010, MNRAS, 408, 1396
\bibitem[Sirressi et al.(2019)]{Sirressi19} Sirressi, M., Cicone, C., Severgnini, P., et al.\ 2019, \mnras, in press (arXiv:1906.00985)
\bibitem[Simpson et al.(1999)]{Simpson99} Simpson, C., Ward, M., O'Brien, P., \& Reeves, J.\ 1999, \mnras, 303, L23 
\bibitem[Tatum et al.(2012)]{Tatum12} Tatum, M.~M., Turner, T.~J., Sim, S.~A., et al.\ 2012, \apj, 752, 94 
\bibitem[Tremaine et al.(2002)]{Tremaine02} Tremaine, S., Gebhardt, K., Bender, R., et al.\ 2002, \apj, 574, 740 
\bibitem[Tombesi et al.(2010)]{Tombesi10} Tombesi, F., Cappi, M., Reeves, J.~N., et al.\ 2010, \aap, 521, A57 
\bibitem[Tombesi et al.(2013)]{Tombesi13} Tombesi, F., Cappi, M., Reeves, J.~N., et al.\ 2013, \mnras, 430, 1102 
\bibitem[Tombesi et al.(2015)]{Tombesi15} Tombesi, F., Mel{\'e}ndez, M., Veilleux, S., et al.\ 2015, \nat, 519, 436 

\bibitem[Torres et al.(1997)]{Torres97} Torres, C.~A.~O., Quast, G.~R., Coziol, R., et al.\ 1997, \apjl, 488, L19 
\bibitem[Veilleux et al.(2013)]{Veilleux13} Veilleux, S., Mel{\'e}ndez, M., Sturm, E., et al.\ 2013, \apj, 776, 27 
\bibitem[Veilleux et al.(2017)]{Veilleux17} Veilleux, S., Bolatto, A., Tombesi, F., et al.\ 2017, \apj, 843, 18 
\bibitem[Vestergaard \& Peterson(2006)]{VestergaardPeterson06} Vestergaard, M., \& Peterson, B.~M.\ 2006, \apj, 641, 689 
\bibitem[Wilkins et al.(2017)]{Wilkins17} Wilkins, D.~R., Gallo, L.~C., Silva, C.~V., et al.\ 2017, \mnras, 471, 4436 

\bibitem[Wilms et al.(2000)]{Wilms00} Wilms, J., Allen, A., \& McCray, R. 2000, ApJ, 542, 914

\bibitem[Zubovas \& King(2012)]{ZubovasKing12} Zubovas, K., \& King, A.\ 2012, \apjl, 745, L34 

\end{thebibliography}
\end{document}